\documentclass[11pt,epsf]{article}
\DeclareMathAlphabet{\scr}{U}{rsfs}{m}{n}
\usepackage{latexsym}
\usepackage{epsfig}
\usepackage[mathscr]{eucal}
\usepackage{amsfonts}
\usepackage{amscd}
\usepackage{cite}
\usepackage{array}   
\usepackage{amssymb}
\usepackage{colordvi}
\usepackage[centertags]{amsmath}
\usepackage{enumerate}
\usepackage{graphicx}
\usepackage{booktabs}
\usepackage{theorem}
\usepackage[footnotesize]{caption}
\usepackage{soul}

\newcommand{\newc}{\newcommand}
\newc{\be}{\begin{equation}}
\newc{\ee}{\end{equation}}
\newc{\bea}{\begin{eqnarray}}
\newc{\eea}{\end{eqnarray}}
\newc{\ol}{\overline}
\newc{\wt}{\widetilde}
\newc{\bs}{\boldsymbol}
\newc{\m}{\mathcal}
\newc{\la}{\langle}
\newc{\ra}{\rangle}

\newcommand{\beq}{\begin{eqnarray}} 
\newcommand{\eeq}{\end{eqnarray}} 
\newcommand{\s}{\smallskip} 
  
\newcommand{\bpmatrix}{\begin{pmatrix}}
\newcommand{\epmatrix}{\end{pmatrix}}
\newcommand{\fr}{\frac}
\newcommand{\diag}{\text{diag}}
\newcommand{\id}{{\rm 1\kern-.12em
\rule{0.3pt}{1.5ex}\raisebox{0.0ex}{\rule{0.1em}{0.3pt}}}}

\renewcommand{\Re}[1]{\text{Re}#1}
\renewcommand{\Im}[1]{\text{Im}#1}

\newcommand{\bc}{\begin{center}}
\newcommand{\ec}{\end{center}}

\newcommand{\ba}{\begin{array}}
\newcommand{\ea}{\end{array}}

\begin{document}
\begin{titlepage}
\vspace*{-2.cm}
\begin{flushright}
{\small
FR-PHENO-2013-017 \\
KA-TP-40-2013 \\
PSI-PR-13-16 \\
SFB/CPP-13-112
}
\end{flushright}
\vspace*{1.5cm}

\begin{center}
{\Large \bf  
\textbf{NMSSMCALC: \\
A Program Package for the Calculation of \\
Loop-Corrected Higgs Boson
Masses \\ and Decay Widths \\ in the (Complex) NMSSM}}
\end{center}
\vskip0.5cm

\begin{center}
J.~Baglio$^{1\,}$, R.~Gr\"ober$^{1\,}$,
M. M\"{u}hlleitner$^{1\,}$, D.T.~Nhung$^{1\,}$, H.~Rzehak$^{2\,}$,
M.~Spira$^{3\,}$,\\ 
J.~Streicher$^{1\,}$ and K. Walz$^{1\,}$
\end{center}

\vskip 20pt

\begin{center}
{\small\it
$^1$Institut f\"ur Theoretische Physik, KIT, D-76128 Karlsruhe,
Germany.}\\[3mm] 
{\small\it
$^2$Albert-Ludwigs-Universit\"at Freiburg, Physikalisches Institut, D-79104 Freiburg, Germany.}\\[3mm]
{\small\it
$^3$Paul Scherrer Institut, CH-5232 Villigen PSI, Switzerland.}\\
\end{center}

\vglue 1.0truecm

\begin{abstract}
\noindent
We present the program package {\tt NMSSMCALC} for the calculation of
the loop-corrected NMSSM Higgs boson masses and decay widths in the
CP-conserving and CP-violating NMSSM. The full one-loop corrections to
the Higgs boson masses are evaluated in a mixed renormalisation scheme of
on-shell and $\overline{\mbox{DR}}$ conditions. The Higgs decay widths
include the dominant higher order QCD corrections, and the decays into
bottom quarks, strange quarks and $\tau$ leptons are supplemented by
higher order SUSY corrections through effective couplings. All
relevant off-shell decays into two massive gauge bosons, gauge and Higgs
boson and Higgs pair final states as well as into heavy quark pairs
are computed.  The input and output files feature the SUSY Les
Houches Accord so that the program can easily be linked with existing
computer tools. 
\end{abstract}
\thispagestyle{empty}
\end{titlepage}
\vfill
\newpage
\setcounter{page}{1}

\section{Introduction}
The announcement of the discovery of a new particle with a mass
of about 126~GeV by the LHC
experiments ATLAS and CMS \cite{:2012gk,:2012gu} has marked a
milestone in the history of particle physics. With the growing amount
of data the experiments have started investigating the properties of
this particle, {\it i.e.}~its couplings to other Standard Model (SM)
particles as well as its spin and parity quantum numbers. The results
of these measurements strongly suggest the particle to be the Higgs boson,
the particle predicted by the Higgs mechanism \cite{Goldstone,Higgs}
which allows to introduce particle masses without violating the
gauge symmetries of the SM. While the data is compatible with a SM-like
Higgs state it leaves room for interpretations within 
Higgs sectors of theories beyond the SM (BSM),
among which supersymmetric (SUSY) 
extensions have been extensively studied. The correct interpretation
of the data and the proper distinction between different models require, on the
theoretical side, precise predictions for the masses and couplings of the
investigated models as well as for the Higgs production cross sections
and branching ratios, taking into account higher order
corrections. The implementation of these calculations in public
computer tools allows to test various models at
the highest possible precision. \s

The Next-to-Minimal Supersymmetric Extension of the SM (NMSSM)
\cite{genNMSSM1,genNMSSM2,Nevzorov:2004ge} 
with two complex Higgs doublets and one complex singlet
field features after electroweak symmetry breaking (EWSB) seven Higgs
bosons. In the CP-conserving case, these are three neutral CP-even, two
neutral CP-odd and two charged Higgs bosons. Allowing for CP violation
the five neutral Higgs mass eigenstates, being mixtures of CP-even and
CP-odd components already at tree level, do not carry definite CP
quantum numbers anymore. 
Besides the benefits of supersymmetric theories in general 
the NMSSM is an attractive BSM extension as it solves the $\mu$
problem of the Minimal Supersymmetric Extension of the SM (MSSM). The
$\mu$-parameter of the Higgs potential is generated dynamically
through the singlet field acquiring 
a vacuum expectation value (VEV) so that it is naturally of the order of the scale
of EWSB. Furthermore, with an additional tree-level contribution to
the lighter MSSM-like Higgs boson mass, proportional to the doublet-singlet
coupling $\lambda$, it requires smaller radiative corrections to
achieve a mass of 126~GeV, and hence smaller stop masses and/or
mixing, implying less fine-tuning \cite{finetune}. The extended Higgs sector
finally, entails a plethora of interesting phenomenological
implications, like Higgs-to-Higgs decays, suppressed or
enhanced branching ratios compared to the SM or a Higgs signal which
is built up by two resonances close in mass around 126~GeV, to cite
only a few of them. This makes clear that precise predictions for
the masses of  
the NMSSM Higgs states and for their production and decay processes
including higher order corrections are needed to interpret
the experimental data reliably. \s

We present the program package {\tt NMSSMCALC}\footnote{A first
  version of the program has been presented in \cite{Baglio:2013vya}.} for the
calculation of loop-corrected NMSSM Higgs boson masses as well as
of the decay widths and branching ratios, both for the CP-conserving and
the CP-violating case. The package includes
\begin{itemize}
\item The computation of the full one-loop corrections to the NMSSM Higgs
  boson masses in a mixed renormalisation scheme of on-shell and
  $\overline{\mbox{DR}}$ conditions both in the CP-conserving 
  \cite{Ender:2011qh} and in the CP-violating NMSSM
  \cite{Graf:2012hh}.\footnote{For further higher order calculations
    to the real NMSSM Higgs boson masses, see
    \cite{effpot,leadlog,Degrassi:2009yq,full1loop}, and for the
    complex case see
    Refs.~\cite{effcorr1,effcorr2,effcorr3,complex2loop,Munir:2013dya}. 
    Corrections 
    to the trilinear Higgs self-couplings have been provided in
    \cite{Nhung:2013lpa}.}
\item The calculation of the Higgs decay widths and branching ratios in the
  CP-conserving and CP-violating implementations of the NMSSM. This
  part of the program package is based on an extension of the Fortran
  code {\tt HDECAY} \cite{hdecay}. The decay widths include the
  dominant higher order QCD corrections. For the neutral Higgs boson decays into a bottom
  quark pair furthermore the higher order SUSY--QCD 
  and the approximate SUSY--electroweak (elw) corrections up to one-loop
  accuracy have been implemented. 
  They have been obtained by adapting the existing
  results for the MSSM 
  \cite{Pierce:1996zz,Carena:1999py,Guasch:2003cv,noth1,noth2,reisser}
  to the NMSSM case.
  The decays into a strange quark pair include the dominant resummed
  SUSY--QCD corrections and the one into a $\tau$ pair the dominant
  resummed SUSY--elw corrections, again by adapting them from
  the MSSM to the NMSSM Higgs bosons. Analogously for the charged Higgs
  boson the higher order SUSY corrections have been implemented for
  the decays into fermion pairs. In the real NMSSM, the decays into
  stop and sbottom pairs, respectively, contain the SUSY--QCD corrections.
\item The inclusion of all relevant off-shell decays into two massive gauge
  boson final states, into gauge and Higgs boson final states, into
  Higgs pairs as well as into heavy quark pairs.
\item The input and output files feature the SUSY Les Houches Accord
  (SLHA) \cite{slha1,slha2} so that the program package can easily be 
 linked with other existing computer tools.
\end{itemize}
In general, the electroweak corrections going beyond the 
approximate SUSY--elw contribution, have not been taken into
account in  the calculation of the decay widths and branching
ratios. They are available for some MSSM decays and could in 
principle be extended to the NMSSM case. Due to the additional NMSSM
singlet field, this requires, however, further
calculations, which are beyond the scope of the present
implementation. We leave this for future work. \s

The outline of the draft is as follows. After introducing the NMSSM
Lagrangian in section~\ref{sec:nmssmlag} we present in  
section~\ref{sec:cpconsloopcorrections} the main features of the
calculation of the one-loop corrections to the NMSSM Higgs boson masses in
the CP-conserving NMSSM, and in
section~\ref{sec:cpviolloopcorrections} in the CP-violating
NMSSM. Section~\ref{sec:decaywidths} gives an overview of the
implemented decays and their higher order corrections. In particular
we present in subsection~\ref{sec:deltab} the higher order SUSY--QCD
and SUSY--elw corrections to the NMSSM Higgs boson decays into
quark pairs. In section~\ref{sec:progdescr} the program
package is described with its main routines, input and output files,
and how to run the program. Section~\ref{sec:slha} deals with various
issues related to the SUSY Les Houches Accord with respect to our
program package. 
In section
\ref{sec:summary} we summarise and give an outlook on future developments.
The program package can be found at the url: {\tt
  http://www.itp.kit.edu/$\sim$maggie/NMSSMCALC/}.

\section{The NMSSM Lagrangian  \label{sec:nmssmlag}} 
With respect to the MSSM Lagrangian, 
the NMSSM differs by the superpotential and the soft SUSY breaking
part. Denoting the Higgs 
doublet superfields, which couple to the up- and down-type quarks, by
$\hat{H}_u$ and $\hat{H}_d$, respectively, and the singlet superfield by
$\hat{S}$, the scale invariant superpotential reads
\beq
W_{NMSSM} = W_{MSSM} - \epsilon_{ij} \lambda \hat{S} \hat{H}^i_d
\hat{H}^j_u + \frac{1}{3} \kappa \hat{S}^3 \;,
\eeq
with the $SU(2)_L$ indices $i,j=1,2$ and the totally antisymmetric
tensor $\epsilon_{12}=1$. While the dimensionless parameters $\lambda$
and $\kappa$ can be complex in general, in case of CP-conservation
they are taken to be real. In terms of the quark and lepton
superfields and their charge 
conjugates, indicated by the superscript $c$, $\hat{Q}, \hat{U}^c,
\hat{D}^c, \hat{L}, \hat{E}^c$, the MSSM superpotential $W_{MSSM}$ is
given by
\beq
W_{MSSM} = \epsilon_{ij} [y_e \hat{H}^i_d \hat{L}^j \hat{E}^c + y_d
\hat{H}_d^i \hat{Q}^j \hat{D}^c - y_u \hat{H}_u^i \hat{Q}^j \hat{U}^c] 
\label{eq:mssmsuperpot} \;,
\eeq
where the colour and generation indices have been suppressed. 
The Yukawa couplings $y_d,$ $y_u$ and $y_e$ in
the MSSM superpotential Eq.~(\ref{eq:mssmsuperpot}) are in
general complex. Their phases can, however, be reabsorbed, by
redefining the quark fields, if generation mixing is neglected as it is
done in our approach for the calculation of the Higgs masses. The MSSM
$\mu$-term as well as the tadpole and bilinear terms of $\hat{S}$ are assumed
to be zero. Denoting the Higgs doublet and singlet component fields by
$H_u$, $H_d$ and $S$, the soft SUSY breaking NMSSM Lagrangian is given by
\beq
\mathcal L_{soft} = {\cal L}_{soft,\, MSSM} - m_S^2 |S|^2 +
(\epsilon_{ij} \lambda 
A_\lambda S H_d^i H_u^j - \frac{1}{3} \kappa
A_\kappa S^3 + h.c.) \;, \label{eq:nmssmsoft}
\eeq
with the soft SUSY breaking MSSM Lagrangian
\beq
{\cal L}_{soft, \, MSSM} &=& -m_{H_d}^2 H_d^\dagger H_d - m_{H_u}^2
H_u^\dagger H_u -
m_Q^2 \tilde{Q}^\dagger \tilde{Q} - m_L^2 \tilde{L}^\dagger \tilde{L}
- m_U^2 \tilde{u}_R^* 
\tilde{u}_R - m_D^2 \tilde{d}_R^* \tilde{d}_R 
\nonumber \\\nonumber
&& - m_E^2 \tilde{e}_R^* \tilde{e}_R - (\epsilon_{ij} [y_e A_e H_d^i
\tilde{L}^j \tilde{e}_R^* + y_d
A_d H_d^i \tilde{Q}^j \tilde{d}_R^* - y_u A_u H_u^i \tilde{Q}^j
\tilde{u}_R^*] + h.c.) \\
&& -\frac{1}{2}(M_1 \tilde{B}\tilde{B} + M_2
\tilde{W}_k\tilde{W}_k + M_3 \tilde{G}\tilde{G} + h.c.) \;.
\label{eq:mssmsoft}
\eeq
Here $\tilde{B}$, $\tilde{W}_k$ ($k=1,2,3$) and $\tilde{G}$ are the
gaugino fields, and $\tilde{Q}=(\tilde{u}_L,\tilde{d}_L)^T$,
$\tilde{L}=(\tilde{\nu}_L,\tilde{e}_L)^T$, where the tilde denotes the 
scalar components of the corresponding quark and lepton
superfields. In the soft SUSY breaking NMSSM Lagrangian
Eq.~(\ref{eq:nmssmsoft}) the soft SUSY breaking mass parameters $m_X^2$
of the scalar fields $X=S,H_d,H_u,Q,U,D,L,E$ are real. The soft
SUSY breaking trilinear couplings $A_x$ ($x=\lambda,\kappa,d,u,e$) and
the gaugino mass parameters $M_1,M_2$ and $M_3$, however, are complex in general but real
in the CP-conserving case. We furthermore neglect squark and
slepton mixing between the generations and set soft SUSY breaking
terms linear and quadratic in the singlet field $S$ to zero. 

\section{Loop-corrected Higgs boson masses in the real
  NMSSM  \label{sec:cpconsloopcorrections}} 
In this section we summarise the main features of the calculation of
the one-loop corrections to the NMSSM Higgs boson masses in the
CP-conserving NMSSM. For details we refer the reader to
Ref.~\cite{Ender:2011qh}.  

\subsection{The CP-conserving NMSSM Higgs sector \label{sec:nmssmhiggssec}}  
The Higgs mass matrix is derived from the NMSSM Higgs potential, which
is obtained from the superpotential, the soft SUSY
breaking terms and from the $D$-term contributions. The Higgs
doublet and singlet fields entering the Higgs potential acquire
non-vanishing vacuum expectation values (VEVs) after electroweak symmetry
breaking. After expansion of
the Higgs fields about their VEVs $v_u$, $v_d$ and $v_s$, chosen to be real
and positive, 
\beq
H_d =
 \bpmatrix (v_d + h_d +ia_d)/{\sqrt 2}\\ h_d^- \epmatrix,\quad
H_u = \bpmatrix
h_u^+ \\ (v_u + h_u +ia_u)/{\sqrt 2}\epmatrix,\quad
S= \fr{v_s + h_s +ia_s}{\sqrt 2} \;,
\label{eq:Higgs_expansion} 
\eeq
the $3\times 3$ Higgs mass matrices squared for the CP-even and CP-odd
component Higgs fields, $M_S^2$ and $M_A^2$, respectively, can be
derived from the tree-level scalar potential. Explicit
expressions can be found in \cite{Ender:2011qh}. The squared mass
matrix $M_S^2$ is diagonalised through a rotation ${\cal
  R}^S$, yielding the CP-even mass eigenstates $H_i$ ($i=1,2,3$),
\beq
 \bpmatrix H_1,H_2,H_3 \epmatrix^T =   {\cal R}^{S} \bpmatrix
 h_d,h_u,h_s \epmatrix^T, 
\quad \diag( (M_{H_1}^{(0)})^2,(M_{H_2}^{(0)})^2,(M_{H_3}^{(0)})^2)=
{\cal R}^S {M_S^2} 
({\cal R}^S)^T \label{eq:Srotation} \; .
\eeq 
The mass eigenstates are ordered by ascending mass, $M_{H_1}^{(0)} \le
M_{H_2}^{(0)} \le M_{H_3}^{(0)}$, where the superscript $(0)$ indicates the tree-level
mass values.  
In the CP-odd Higgs sector a first rotation ${\cal R}^G$ is applied to
separate the massless Goldstone boson $G$, followed by a 
rotation ${\cal R}^P$ to obtain the mass eigenstates $A_i \equiv A_1,
A_2, G$ ($i=1,2,3$), {\it cf.}~\cite{Ender:2011qh},
\beq
 \bpmatrix A_1,A_2,G \epmatrix^T =   {\cal R}^{P}  {\cal R}^G \bpmatrix
 a_d,a_u,a_s \epmatrix^T, 
\quad \diag((M_{A_1}^{(0)})^2,(M_{A_2}^{(0)})^2,0)= {\cal R}^P {\cal
  R}^G {M_A^2} 
({\cal R}^P {\cal R}^G)^T \label{eq:Protation} \; .
\eeq 
At tree-level the CP-conserving NMSSM Higgs potential depends on 12
independent parameters, which are the $U(1)$ and $SU(2)$ gauge
couplings $g_1$ and $g_2$,
the three VEVs,  the soft SUSY breaking mass parameters of the doublet
and the singlet Higgs fields and the NMSSM specific parameters and
soft SUSY breaking couplings,
\beq 
g_1,g_2,v_d,v_u,v_s,m_{H_d}^2,m_{H_u}^2,m_S^2,\lambda, \kappa,A_\lambda,
A_\kappa  \;. 
\eeq
Some of these parameters are replaced in order to obtain a more
transparent physical interpretation.  The minimisation conditions of
the Higgs potential $V$ require the terms linear in the Higgs fields
to vanish in the vacuum. Hence, at tree-level, for the scalar fields,
\beq
t_\phi \equiv \left. \frac{\partial V}{\partial \phi}
\right|_{\mbox{\scriptsize Min}} \stackrel{!}{=} 0 \qquad \mbox{for}
\qquad 
\phi = h_d, h_u, h_s \;.
\eeq
These conditions are used to trade $m_{H_d}^2$, $m_{H_u}^2$ and
$m_S^2$ for the tadpole parameters $t_{h_d}$, $t_{h_u}$ and
$t_{h_s}$. The soft SUSY breaking 
coupling $A_\lambda$ can optionally be replaced by the charged Higgs boson
mass $M_{H^\pm}$, and the parameters $g_1, g_2, v_u, v_d$ are
substituted by the gauge boson
masses $M_W$ and $M_Z$, the electric charge $e$ and $\tan\beta$, where
$\tan\beta$ is the ratio of the two vacuum expectation values of the
doublet fields,
\beq
\tan\beta = \frac{v_u}{v_d} \;.
\eeq 
We hence have two possible new parameter sets to work with 
\bea
\text{1st parameter set: }& &M_Z,M_W, M_{H^\pm},t_{h_d},t_{h_u},t_{h_s},e,\tan\beta,\lambda,
v_s,\kappa, A_\kappa\; ;\label{eq:1parset} \\ 
\text{2nd parameter set: }& &M_Z,M_W,t_{h_d},t_{h_u},t_{h_s},e,\tan\beta,\lambda,A_\lambda,
v_s,\kappa, A_\kappa\;. \label{eq:2parset}
\eea
Note that the first one is the one we chose in our previous work
\cite{Ender:2011qh}, whereas the second one provides an additional option. 

\subsection{One-loop corrected Higgs boson masses in the real NMSSM  \label{sec:oneloopmass}}
The Higgs self-energies calculated for the determination of the
loop-corrected Higgs boson masses develop ultraviolet (UV)
divergencies. The renormalisation of the parameters entering the loop
calculation renders the result finite. The two renormalisation schemes
which we apply to the two parameter sets are a mixture of on-shell and
$\overline{\mbox{DR}}$ renormalisation conditions as follows:
\bea
\text{1st renormalisation scheme: }& &\underbrace{M_Z,M_W,
  M_{H^\pm},t_{h_d},t_{h_u},t_{h_s},e}_{\mbox{on-shell
    scheme}},\underbrace{\tan\beta,\lambda, v_s,\kappa,
  A_\kappa}_{\overline{\mbox{DR}} \mbox{ scheme}} \;; \nonumber \\
  \text{2nd renormalisation scheme: }& &\underbrace{M_Z,M_W,
  t_{h_d},t_{h_u},t_{h_s},e}_{\mbox{on-shell
    scheme}},\underbrace{\tan\beta,\lambda,A_\lambda, v_s,\kappa,
  A_\kappa}_{\overline{\mbox{DR}} \mbox{ scheme}} \;. \label{eq:renscheme}
\eea
The specific form of the counterterms for the first scheme can be found in
\cite{Ender:2011qh}. The only modification in the second scheme is that
the counterterm to $A_\lambda$ is determined from the 
charged Higgs sector via a $\overline{\mbox{DR}}$ condition.
In both schemes the chargino sector is
exploited in the derivation of the counterterm related to $v_s$, and
the neutralino sector in the derivation of the counterterm for
$\kappa$. The field renormalisation of the Higgs boson doublet and
singlet fields are defined via $\overline{\mbox{DR}}$
conditions. External self-energy contributions are properly taken into
account, {\it cf.}~\cite{Ender:2011qh,Frank:2006yh} for details.  
\s

The parameters given in the sets Eqs.~(\ref{eq:1parset}) and
(\ref{eq:2parset}) are the ones which we
use in the calculation of the mass corrections and on which we apply
our renormalisation conditions. However, they are not the input
parameters, which are to be provided by the user. These are the SM
inputs as defined in the SLHA\footnote{They are given by the inverse
  electromagnetic coupling and the strong
  coupling at the $Z$ pole in the
  $\overline{\mbox{MS}}$ scheme with five active flavours, the Fermi
  constant, the $Z$ pole mass, the running mass of the $b$-quark in the
  $\overline{\mbox{MS}}$ scheme and the top and tau pole masses.
Additionally the $W$ boson pole mass has to be given. This value is not
included in the original SLHA SM inputs, but needed in the calculation
of the loop-corrected Higgs boson masses and of the decay widths.}, 
the value for $\tan\beta$, the soft SUSY breaking gaugino and squark
mass parameters, the soft SUSY breaking trilinear couplings and the NMSSM
specific parameters $\lambda$, $\kappa$, $A_\lambda$, $A_\kappa$ and
$\mu_{\scriptsize \mathrm{eff}}$. The latter is  
generated dynamically when the Higgs singlet field $S$ acquires its
vacuum expectation value $\langle S \rangle$,
\beq
\mu_{\scriptsize \mathrm{eff}} = \lambda \langle S \rangle \equiv
\lambda \, \frac{v_s}{\sqrt{2}}\;.
\eeq  
Alternatively, for $A_\lambda$ the charged Higgs boson mass $M_{H^\pm}$ can be provided. \s

The one-loop corrected scalar Higgs boson masses squared are determined
numerically and given by the real parts of the zeroes of the determinant of the
two-point vertex functions $\hat{\Gamma}^S$, 
\beq
\hat{\Gamma}^S (k^2) = && \phantom{LLLLLLLLLLLLLLLLLLLLLLLLLLLLLLLLLLLLLLLLLLLLLLLLLLLL} \nonumber\\
&& \hspace*{-2cm} i \left( \begin{array}{ccc} k^2 - (M_{H_1}^{(0)})^2 +
    \hat{\Sigma}_{H_1 H_1} (k^2) & \hat{\Sigma}_{H_1 H_2} (k^2) &
    \hat{\Sigma}_{H_1 H_3} (k^2) \\
\hat{\Sigma}_{H_2 H_1} (k^2) & k^2 - (M_{H_2}^{(0)})^2 +
\hat{\Sigma}_{H_2 H_2} (k^2) & \hat{\Sigma}_{H_2 H_3} (k^2) \\
\hat{\Sigma}_{H_3 H_1} (k^2) & \hat{\Sigma}_{H_3 H_2} (k^2) &
k^2 - (M_{H_3}^{(0)})^2 +\hat{\Sigma}_{H_3 H_3} (k^2)
\end{array} \right)  
\hspace*{-2cm }\label{eq:cpevenoneloop}
\eeq
while the pseudoscalar masses squared are extracted from
$\hat{\Gamma}^P$, 
\beq
 \hat{\Gamma}^P (k^2) =  i \left( \begin{array}{cc} k^2 - (M_{A_1}^{(0)})^2 +
    \hat{\Sigma}_{A_1 A_1} (k^2) & \hat{\Sigma}_{A_1 A_2} (k^2) \\
\hat{\Sigma}_{A_2 A_1} (k^2) & k^2 - (M_{A_2}^{(0)})^2 +
\hat{\Sigma}_{A_2 A_2} (k^2) 
\end{array} \right) \;. 
\label{eq:cpoddoneloop} 
\eeq
Note that in the pseudoscalar sector no mixing with the would-be
Goldstone bosons is taken into account, since we checked that the
effect is numerically negligible. Also no mixing
with the longitudinal component of the $Z$ boson is taken into
account, as it has been shown in the MSSM \cite{Hollik:2002mv} that it
is sufficient to include the mixing with the would-be Goldstone
boson. The unrenormalised self-energies and tadpole conditions appearing
implicitly in the renormalised self-energies $\hat{\Sigma}$ are
evaluated at one-loop order. The mass eigenvalues are obtained in an
iterative procedure keeping the full dependence on the external
momentum squared $k^2$ in the renormalised self-energies. The Higgs
mixing matrix elements on the other hand are obtained by setting
$k^2=0$, corresponding to a proper definition of the effective mixing
matrix elements. This
yields a unitary mixing matrix.\footnote{For non-vanishing $k^2$ the mixing
  matrix is not unitary.} The differences between this approach
and the one with non-vanishing $k^2$ have been found to be small in the
investigated parameter sets.  \s

In the first renormalisation scheme the mass of the charged
Higgs boson is renormalised on-shell, and therefore constitutes the
physical value even at one-loop order. If, however, 
the second scheme is applied the pole mass of the charged Higgs boson
at one-loop level, $M_{H^\pm}^{\text{1loop}}$, is determined
by iteratively solving the equation
\beq
(M_{H^\pm}^{\text{1loop}})^2=(M_{H^\pm}^{\text{tree}})^2-
\hat{\Sigma}_{H^\pm H^\pm}\big((M_{H^\pm}^{\text{1loop}})^2\big)\;,
\eeq
with $\hat\Sigma_{H^\pm H^\pm}$ denoting the renormalised self-energy 
of the charged Higgs boson which is given by
\beq
\hat\Sigma_{H^\pm H^\pm}=
\Sigma_{H^{\pm}H^{\pm}}(k^2)+(k^2-M^2_{H^{\pm}})(s_\beta^2  
\delta Z_{H_d} 
                 +c_\beta^2 \delta Z_{H_u})-\delta M_{H^{\pm}}^2 \;,
\eeq 
where the counterterm $\delta M_{H^\pm}^ 2$ is a function of the
counterterms of the second renormalisation scheme and the wavefunction
renormalisation factors are renormalised in the $\overline{\mbox{DR}}$
scheme. Again, external self-energy contributions are properly taken into
account~\cite{Ender:2011qh}. The mixing of the charged Higgs boson with
the charged would-be Goldstone boson is neglected. \s 

Throughout the calculation of the one-loop masses 
the running $\overline{\mbox{DR}}$ top and bottom quark masses are used.

\section{Loop-corrected Higgs boson masses in the complex
  NMSSM  \label{sec:cpviolloopcorrections}} 
The calculation of the Higgs boson masses in the CP-violating NMSSM is a
generalization of the real case described in section
\ref{sec:oneloopmass}. Choosing 
vanishing phases and imaginary parts, respectively, yields the same result as
before. Hence, in this section we restrict ourselves to summarising the
differences of 
the CP-violating NMSSM with respect to the CP-conserving NMSSM
described in the previous section. For further details, we refer the
reader to Ref.~\cite{Graf:2012hh}. \s

\subsection{The CP-violating NMSSM Higgs sector}
In addition to the complex parameters present in the MSSM, 
there are four more parameters in the NMSSM, that can become
complex. These are the couplings $\lambda$, $\kappa$ and the soft SUSY
breaking trilinear couplings $A_\lambda$ 
and $A_\kappa$. The complex parameters originating from the MSSM part
are the soft SUSY breaking trilinear couplings $A_u$, $A_d$, $A_e$ of
the up-type, down-type and charged lepton-type sfermions,
respectively, as well as the soft SUSY breaking mass parameters of the
gauginos $M_1$, $M_2$ and $M_3$. $R$-symmetry can be
exploited to choose either $M_1$ or $M_2$ to be real. We do not make
use of this symmetry, however, in order to keep the relations as
general as possible. The kinetic and
gauge interaction parts of the NMSSM Lagrangian contain no complex
parameters. \s

In the expansion of the Higgs boson fields about the VEVs two further
phases, $\varphi_u$ and $\varphi_s$, appear,
\begin{align}
H_d = \begin{pmatrix}\frac{1}{\sqrt{2}}(v_d + h_d + i a_d) \\ 
                      h_d^- \end{pmatrix}~, \;\;\; 
H_u = e^{i \varphi_u} \begin{pmatrix} h_u^+ \\
            \frac{1}{\sqrt{2}}(v_u + h_u + i a_u) \end{pmatrix}~, \;\;\; 
S = \frac{e^{i \varphi_s}}{\sqrt{2}} (v_s + h_s + i a_s)~.
\label{Higgsdecomp}
\end{align}
They describe the phase differences between the three VEVs $\langle
H_d^0 \rangle$, $\langle H_u^0 \rangle$ and $\langle S \rangle$. For
phase values $\varphi_u=\varphi_s= n\pi$, $n \in \mathbb{N}$, the fields $h_i$ and $a_i$
($i=d,u,s$) are the pure CP-even and CP-odd parts of the neutral entries of
$H_u$, $H_d$ and $S$. We exploit the freedom in the phase choice of
the Yukawa couplings to set $\varphi_{y_u}= - \varphi_u$ and 
assume the down-type and charged lepton-type Yukawa couplings to be
real. In this way the quark and lepton mass terms yield real masses
without any further phase transformation of the corresponding
fields. \s

After the expansion about the VEVs the Higgs boson mass matrix
$M_{\phi\phi}$ can be read off from the terms in the Higgs potential
which are bilinear in the neutral Higgs boson fields. CP-violation
introduces a mixing between CP-even and CP-odd component 
fields, so that this matrix is a $6\times 6$ matrix in the basis $\phi
= (h_d, h_u, h_s, a_d, a_u, a_s)^T$, which can be expressed in terms of three $3
\times 3$ matrices $M_{hh}, M_{aa}$ and $M_{ha}$,
\begin{align}
M_{\phi\phi} = \begin{pmatrix} M_{hh}&  M_{ha}\\
                          M_{ha}^T &
                          M_{aa}\end{pmatrix} \;,
\label{eq:higgsmassmatrix}
\end{align}
where $M_{hh}$ and $M_{aa}$ are symmetric matrices, describing the
mixing among the CP-even components of 
the Higgs doublet and singlet fields and among the CP-odd components,
respectively.
In case of CP-conservation the 
matrix $M_{ha}$, which mixes CP-even and CP-odd components,
vanishes. Note that due to the application of the 
minimisation conditions of the Higgs potential $V$,
\beq
t_\phi \equiv \left. \frac{\partial V}{\partial \phi}
\right|_{\mbox{\scriptsize Min}} \stackrel{!}{=} 0 \qquad \mbox{for}
\qquad \phi = 
h_d, h_u, h_s, a_d, a_u, a_s \;,
\eeq
in the tree-level Higgs sector only one linearly independent phase combination
$\varphi_y$ appears after applying the tadpole conditions for $\phi=a_d$
and $a_s$\footnote{The tadpole condition for $a_u$ does not lead to a
  new linearly independent condition. Note, that in the real case, the
  CP-odd tadpole conditions vanish and are thus 
  automatically fulfilled.},  
\beq
\varphi_y = \varphi_\kappa - \varphi_\lambda + 2\varphi_s - \varphi_u \;.
\eeq
Hence the CP mixing due to $M_{ha}$ is governed by $\sin\varphi_y$ at
tree-level. \s

The transformation into mass eigenstates is performed by subsequently
applying the $6 \times 6$ rotation matrix ${\cal R}^G$ to separate the
would-be Goldstone boson field and then the matrix ${\cal R}$ to
rotate to the mass eigenstates $H_i$ ($i=1,...,5$), yielding a
diagonal mass matrix squared,
\beq
(H_1,H_2,H_3,H_4,H_5,G)^T &=& {\cal R}^D 
(h_d,h_u,h_s,a_d,a_u,a_s)^T \label{eq:cpR}\\
\mbox{diag} ((M_{H_1}^{(0)})^2,...,(M_{H_5}^{(0)})^2,0) &=& {\cal R}^D 
M_{\phi\phi} ({\cal R}^D)^T \;, \label{eq:cpRm}
\eeq
with ${\cal R}^D \equiv {\cal R} {\cal R}^G$ and the superscript $(0)$
indicating tree-level masses. \s 

At tree-level the parameters which enter the Higgs potential of the
CP-violating NMSSM are
\begin{align}
m_{H_d}^2, m_{H_u}^2, m_S^2, g_1, g_2, v_u, v_d,  v_s, \varphi_s, \varphi_u, 
 \Re\lambda, \Im\lambda,\Re A_{\lambda},\Im A_{\lambda}, \Re\kappa,
\Im\kappa, \Re A_{\kappa},\Im A_{\kappa}~.\label{eq:orgparset}
\end{align}
Again we trade some of the parameters for more physical ones and use
the following two parameter sets
\begin{align}
\underbrace{t_{h_d}, t_{h_u}, t_{h_s}, t_{a_d}, t_{a_s}, M_{H^\pm}^2, M_W^2, M_Z^2, e}_{\mbox{on-shell}}, 
\underbrace{ \tan \beta,   v_s, \varphi_s, \varphi_u, \Re\lambda,
  \Im\lambda, \Re\kappa,  
\Im\kappa, \Re A_{\kappa}}_{\overline{\mbox{DR}}} \;;
\label{eq:defparset}
\end{align}
\begin{align}
\underbrace{t_{h_d}, t_{h_u}, t_{h_s}, t_{a_d}, t_{a_s}, M_W^2, M_Z^2, e}_{\mbox{on-shell}}, 
\underbrace{ \tan \beta,   v_s, \varphi_s, \varphi_u, \Re\lambda,
  \Im\lambda,\Re A_\lambda, \Re\kappa,  
\Im\kappa, \Re A_{\kappa}}_{\overline{\mbox{DR}}} \;.
\label{eq:defparsetb}
\end{align}
The first part of parameters is defined via on-shell conditions and thus
related to physical quantities. This also holds for the tadpole
parameters, as their introduction is motivated by physical
interpretation. The remaining parameters are interpreted as
$\overline{\mbox{DR}}$ parameters. Note that instead of splitting 
the complex parameters $\lambda$, $\kappa$, $A_\lambda$ and $A_\kappa$
into their absolute values and phases, as we did in our previous
work \cite{Graf:2012hh}, we split them into their real and imaginary
parts. This allows us to be in accordance with the SLHA conventions
which require the real part of $A_\lambda$ and $A_\kappa$ to be
$\overline{\mbox{DR}}$ parameters. Expressing the imaginary parts of
$A_\lambda$ and $A_\kappa$ 
by the tadpole parameters $t_{a_d}$ and $t_{a_s}$ then leads to a 
slightly modified renormalisation scheme\footnote{This change in the
  renormalisation scheme does not lead to any significant changes for the loop corrected mass values.}. The two renormalisation 
schemes we provide based on the two parameter sets given above 
differ by either using the mass of the charged Higgs boson as an
on-shell input or the real part of $A_\lambda$ as a $\overline{\mbox{DR}}$
parameter. \s

Again the parameters given in the sets Eqs.~(\ref{eq:defparset})
and (\ref{eq:defparsetb}) are the ones used in
the calculation of the mass corrections and on which the 
renormalisation conditions are applied. The input
parameters provided by the user, however, are the same as the ones for
the real NMSSM, supplemented in the complex NMSSM by the imaginary
parts of the soft SUSY breaking gaugino mass parameters, 
of the soft SUSY breaking trilinear couplings and of $\lambda$,
$\kappa$, $\mu_{\scriptsize \mathrm{eff}}$ as well as the phase
$\varphi_u$, depending on the CP-violating scenario under investigation. The
imaginary parts of $A_\lambda$ and $A_\kappa$ are already fixed by
the tadpole conditions. Note that in the complex case the effective
$\mu$-parameter $\mu_{\scriptsize \mathrm{eff}}$ is defined as
\beq
\mu_{\scriptsize \mathrm{eff}} = \lambda \langle S \rangle \equiv
\lambda \,
\frac{v_s e^{i\varphi_s}}{\sqrt{2}}\;.
\eeq

\subsection{One-loop corrected NMSSM Higgs boson masses in
 the complex NMSSM \label{sec:complexoneloop}}
The one-loop corrected neutral Higgs boson masses squared are
determined numerically as the real parts of the zeroes of the determinant of the
two-point vertex functions $\hat{\Gamma}$,
\beq
\hat\Gamma(k^2)=i\big(\id\cdot k^2-\mathcal{M}^{\text{1l}}\big)\qquad \text{with} \qquad 
\big(\mathcal{M}^{\text{1l}}\big)_{ij}= \big(M^{(0)}_{H_i}\big)^2\delta_{ij}-\hat\Sigma_{ij}(k^2)\quad i,\,j = 1,\dots,5~,
\label{eq:sematrix}
\eeq
where the superscript $\text{1l}$ denotes the one-loop order. The specific
form of the renormalised self-energies $\hat{\Sigma}_{ij}$ in terms of
the 1-loop self-energies, field renormalisation matrices and
counterterms, to render the self-energies finite, can be found in
\cite{Graf:2012hh}. This reference includes the detailed description of
the field renormalisation 
and of the renormalisation of the parameters given in
Eq.~(\ref{eq:defparset}). We make use of the chargino and
neutralino sector in order to determine the counterterms of $v_s,
\varphi_s, \lambda, \kappa$ and $\varphi_u$ as well as those of $M_1$ and
$M_2$. \s

As in the CP-conserving case the one-loop Higgs masses are obtained
iteratively by keeping the dependence on the external momentum
squared in the renormalised self-energies. 
The mixing elements of the matrix performing the rotation from the
interaction to the mass eigenstates at one-loop level are extracted
for zero external momenta, {\it i.e.}~the rotation matrix relating the
tree-level to the one-loop mass eigenstates is defined as the matrix
which diagonalises ${\cal M}^{\text{1l}}$ after setting the momenta in
${\cal M}^{\text{1l}}$ to zero. While this procedure does not retain
the full momentum dependence it has the advantage of yielding a mixing
matrix which is unitary. We have checked
numerically that the difference with respect 
to the result keeping the full momentum dependence is negligible.
Once again we use the running top and bottom quark masses, and the charged
Higgs boson mass is obtained as before in the CP-conserving case.

\section{Decay Widths \label{sec:decaywidths}}
The spectrum file with the loop-corrected NMSSM Higgs masses and
mixings and all SUSY particles with corresponding mixing angles is
used in the routine which calculates the decay widths. The included decays are
the NMSSM Higgs boson decays into SM and SUSY particle pairs as well as
off-shell decays into 3- or 4-particle final states. They are described
in more detail in the following. 

\subsection{Decay Widths in the CP-conserving
  NMSSM \label{sec:decaywidthsreal}}
\noindent
\underline{Decays into quark pairs:}
The neutral Higgs decay widths into quark pairs receive QCD
corrections which are 
available for the SM including the fully massive next-to-leading order
(NLO) corrections near
threshold \cite{nearthreshold} and massless ${\cal O} (\alpha_s^4)$
corrections far above threshold
\cite{abovethreshold,chetyrkin,baikov}. Furthermore large logarithms
are resummed through the running of the quark masses and the strong
coupling constant. The QCD corrections for the charged Higgs boson
decay into a heavy quark pair have been given in \cite{Djouadi:1994gf}. 
As these QCD corrections factorise with respect to
the tree-level amplitude they can be taken over for the NMSSM Higgs decays
and have been included in the decay code. In the decays into bottom
quarks, apart from the QCD 
corrections, higher order SUSY corrections have been included by
absorbing them into effective Yukawa couplings. They include the
resummation of the dominant corrections for large values of
$\tan\beta$ \cite{Pierce:1996zz,Carena:1999py,Guasch:2003cv} and the
SUSY--QCD corrections to the leading SUSY--QCD and 
top-induced SUSY--elw contributions
\cite{noth1,noth2,reisser}. We have 
adapted these results for the MSSM Higgs bosons to the NMSSM case.
Furthermore, the resummed corrections have been included in
the decays into a $\tau$ pair and into a strange quark pair. Details on
the determination of these corrections are given in section
\ref{sec:deltab}. \s

For the decays of the heavier neutral Higgs bosons into a top quark
pair below the threshold off-shell decays can be sizable and have been
included in the program. For the charged Higgs boson, off-shell
decays below the top-bottom, the top-strange and the top-down quark
threshold, respectively, have been taken into account. The decay
widths have been obtained from \cite{offshellphi} 
by making the necessary changes for the NMSSM case. In the charged
Higgs boson decays into quarks we have taken into account generation
mixing through the Cabibbo--Kobayashi--Maskawa (CKM) mixing matrix elements. \s

\noindent \underline{Decays into gluons:}
The decay width of a neutral Higgs boson into gluons, a
loop-induced process already at tree-level, is mediated by quark loops
and, in case of CP-even Higgs bosons, in addition by squark
loops. The QCD  corrections, which can be taken over from the SM,
respectively, MSSM case, have been included up to N$^3$LO in the limit
of heavy quark
\cite{Chetyrkin:1997un,Kramer:1996iq,Schroder:2005hy,Chetyrkin:2005ia,nloggqcd,Spira:1993bb,Spira:1995rr,Baikov:2006ch}
and squark \cite{Dawson:1996xz}\footnote{For the NLO QCD corrections
  including the full mass dependence, see
  \cite{Spira:1993bb,Spira:1995rr}, and for the 
  (SUSY--)QCD corrections including the full squark mass dependence,
  see
  \cite{Harlander:2003bb,Anastasiou:2006hc,Muhlleitner:2006wx,Anastasiou:2008rm,Degrassi:2008zj,Muhlleitner:2008yw,Degrassi:2010eu}.}
loop particle masses. The electroweak corrections are unknown for
supersymmetric Higgs boson decays.\s 

\noindent \underline{Decays into a pair of photons:}
Also the decay into a photon pair is loop-mediated already at lowest
order, including $W$ boson, heavy fermion, charged Higgs boson, sfermion
and chargino loops for the scalar Higgs boson decays and heavy fermion
and chargino loops for the pseudoscalar ones \cite{Spira:1995rr}. The
QCD corrections to quark and squark loops have been calculated
including the full mass dependence both for the quarks
\cite{Spira:1995rr,hgagaqcd}\footnote{Threshold effects have been
  discussed in \cite{Melnikov:1994jb}.} and squarks 
\cite{Muhlleitner:2006wx}. These corrections can be taken over to the
NMSSM and have been included in the program. The genuine
SUSY--elw corrections for photonic SUSY Higgs
decays are unknown. \s 

\noindent \underline{Decays into $Z\gamma$:}
The loop induced decays of scalar Higgs bosons into $Z\gamma$ are
mediated by $W$, heavy fermion, charged Higgs, sfermion and chargino
loops, while the pseudoscalar decays proceed only through charged
fermion and chargino loops. While the electroweak corrections are
unknown, the QCD corrections to quark loops are numerically small
\cite{nloZga} and have not been taken into account. \s

\noindent \underline{Decays into massive gauge bosons:}
The decay width of the scalar NMSSM Higgs bosons into massive gauge
bosons can be obtained from the SM decay width by replacing the SM
Higgs coupling to gauge bosons by the corresponding NMSSM Higgs
coupling. We have included the option of double off-shell
decays \cite{cahn} in the program. Electroweak corrections to the decay
have not been calculated for the NMSSM case and are therefore not
included. The pseudoscalar Higgs bosons do not decay into massive
gauge bosons at tree-level. \s

\noindent \underline{Decays into Higgs bosons:}
The heavier Higgs particles can decay into a pair of lighter Higgs
bosons. Due to the enlarged Higgs sector various
Higgs-to-Higgs decays are possible depending on the mass hierarchies
\cite{Miller:2003ay}. The following decays have been included in the program
($j,k=1,2,l=1,2,3$), 
\beq
\begin{array}{lllllllll}
H_1 &\to& A_1 A_1 \,, &\quad H_1 &\to& A_1 A_2  \\
H_2 &\to& H_1 H_1\,, &\quad H_2 &\to& A_j A_k  \\ 
H_3 &\to& H_j H_k\,, &\quad H_3 &\to& A_j A_k \\
A_2 &\to& A_1 H_l\,. & &&& 
\end{array}
\eeq
The contributions from off-shell final
states may be significant. We have therefore included double off-shell
decays into two Higgs bosons with the Higgs bosons subsequently
decaying into fermions. For $M_{H_i} > M_{H_j}$ and $M_{H_i} >
M_{H_k}$ where $H_{i,j,k}$ denotes generically scalar and
pseudoscalar Higgs bosons, the decay width is given by
\beq
\Gamma (H_i \to H_j^* H_k^*) &=& \frac{1}{\pi^2} \int_0^{M_{H_i}^2}
\frac{dQ_1^2 M_{H_j} \Gamma_{H_j}}{(Q_1^2-M_{H_j}^2)^2+M_{H_j}^2
  \Gamma_{H_j}^2} \int_0^{(M_{H_i}-Q_1)^2}  \frac{dQ_2^2 M_{H_k}
  \Gamma_{H_k}}{(Q_2^2-M_{H_k}^2)^2+M_{H_k}^2 
  \Gamma_{H_k}^2}  \Gamma_0 \nonumber \\
\hspace*{-1cm} \mbox{with} && \\
\Gamma_0 &=& \lambda_{H_i H_j H_k}^2 \delta_{H} \frac{G_F M_Z^4}{16
  \sqrt{2} \pi M_H} \lambda (Q_1^2,Q_2^2;M_{H_i}^2) \frac{Q_1^2
  Q_2^2}{M_{H_j}^2 M_{H_k}^2} \;,
\eeq
where $\lambda_{H_i H_j H_k}$ denotes the trilinear Higgs coupling
normalised to $(\sqrt{2} G_F)^{1/2} M_Z^2$,
$\delta_H =2$ for $H_j \ne H_k$ 
and with the two-body phase space function
\beq
\lambda (x,y,;z) = \sqrt{(1-x/z -y/z)^2 - 4xy/z^2} \; .  \label{eq:2body}
\eeq

\noindent \underline{Decays into a gauge and a Higgs boson:}
The Higgs bosons can also decay into a gauge and a Higgs boson, where
in the NMSSM there is a plethora of possible decays, given by ($j=1,2,l=1,2,3$)
\beq
\begin{array}{lllllllll}
H_1 &\to& A_1 Z \,, &\quad A_1 &\to& H_j Z  \\
H_{2,3} &\to& A_j Z \,, &\quad H_{2,3} &\to& H^\pm W^\mp  \\ 
A_2 &\to& H_l Z\,, &\quad A_2 &\to& H^\pm W^\mp \\
H^\pm &\to& H_l W^\pm \,, &\quad H^\pm &\to& A_j W^\pm \;.
\end{array}
\eeq
The formulae for the decay widths can be easily obtained from the MSSM
results (see {\it e.g.}~\cite{fortsch}) with the corresponding
replacements of the involved couplings. These decays have been
implemented in the program as well as the decays into a Higgs boson
and an off-shell gauge boson which can be important. They have been
obtained by adapting the MSSM formulae \cite{offshellphi} accordingly.
\s

\noindent \underline{Decays into SUSY particles:}
The decays into chargino or neutralino pairs
\cite{Djouadi:1996pj,Djouadi:1996mj} can reach branching 
ratios of up to 100\% if they are kinematically allowed. They have
been included in the program by adapting the MSSM Higgs couplings to
neutralinos/charginos appropriately. Also the decays into sfermions of
the third generation can become important if kinematically allowed
\cite{Djouadi:1996pj}. We have included them for all generations. The
SUSY--QCD corrections to the stop and sbottom decays of the MSSM Higgs
bosons have been provided in \cite{Bartl:1997yd,Eberl:1999he} and
reanalysed in \cite{Accomando:2011jy}. We have adapted them
to the NMSSM decays and included them in the program. 

\subsection{Decay Widths in the CP-violating
  NMSSM \label{sec:decaywidthscomplex}}
In this section we list the changes in the Higgs boson decays for the 
CP-violating NMSSM. We start with the neutral Higgs bosons and give
explicit formulae for the loop-induced decays into gluons and into photons. \s

\noindent \underline{Decays into gluons:}
We introduce the Feynman rules for the neutral CP-violating Higgs
bosons $H_i$ ($i=1,...,5)$ to fermions as
\beq
-\frac{igm_f}{2 M_W} \left[g_{H_iff}^S -i\gamma_5 \, g_{H_iff}^P\right] \;,
\eeq
with the real coupling factors for up- and down-type fermions,
respectively, 
\beq
g_{H_i ff}^S &=& \left\{ \begin{array}{ll} \frac{{\cal R}_{i2}}{\sin\beta} & \;
    \mbox{for } f = \mbox{ up-type} \\
\frac{{\cal R}_{i1}}{\cos\beta} & \; \mbox{for } f = \mbox{
  down-type} \end{array} \right.  
\eeq
and
\beq
g_{H_i ff}^P &=& \left\{ \begin{array}{ll} \frac{{\cal R}_{i4}}{\tan\beta} & \;
    \mbox{for } f = \mbox{ up-type} \\
{\cal R}_{i4}\tan\beta & \; \mbox{for } f = \mbox{
  down-type} \end{array} \right.  \; .
\eeq
Here ${\cal R}_{ij}$ ($i,j=1,...,5$) denote the elements of the
matrix, which rotates the interaction states to the Higgs mass
eigenstates as defined in
Eqs.~(\ref{eq:cpR}) and (\ref{eq:cpRm}). In the gluonic decay width only the
diagonal Higgs coupling to sfermions $\tilde{f}_{j}$ ($j=1,2$)
appears, which is defined as
\beq
- ig \frac{M_Z^2}{M_W} g_{H_i \tilde{f}_j \tilde{f}_j} \;.
\eeq
Note that the coupling factor $g_{H_i \tilde{f}_j \tilde{f}_k}$ for
the coupling to two different sfermions ($j\ne k$) can be
complex in the CP-violating NMSSM. As the expression for $g_{H_i
  \tilde{f}_j \tilde{f}_j}$ is quite lengthy we refer the reader to
the program code, where it can be found explicitly. With these
coupling definitions the tree-level decay width into 
gluons can be cast into the form
\beq
\Gamma(H_i \to gg) = \frac{G_F \alpha_s^2 M_{H_i}^3}{64 \sqrt{2}\pi^3
    } && \hspace*{-0.7cm} \left( \left|
      \sum_{q}  g_{H_i qq}^S A^S_{1/2}(\tau_q) +
      \sum_{\tilde{q}} \frac{M_Z^2}{m_{\tilde{q}}^2} g_{H_i \tilde{q} \tilde{q}} 
        A_0(\tau_{\tilde{q}})
      \right|^2 \right. \nonumber \\
&& \hspace*{-0.6cm} \left. + 4 \left| \sum_{q}  g_{H_i qq}^P A^P_{1/2}(\tau_q)
      \right|^2\, \right) \;,
\eeq
with $\tau_x = 4 m_x^2/M_{H_i}^2$ ($x=q,\tilde{q}$) and the loop
functions \cite{loggres}
\begin{align}
  A_{1/2}^S(\tau) = \,\, & 2 \tau \Big[
  1+\left(1-\tau\right)f(\tau)\Big]\nonumber\\
  A_{1/2}^P(\tau) = \,\, & \tau f(\tau)\nonumber\\
  A_{0}(\tau) = \,\, & - \tau \Big[1-\tau
  f(\tau)\Big]\label{3loopfunctions} \;,
\end{align}
where
\beq
f (\tau) = \left\{\begin{matrix}\displaystyle
    \arcsin^2\left(\frac{1}{\sqrt{\tau}}\right) & \tau \geq 1\\ 
    \displaystyle -\frac14
    \left[\ln\left(\frac{1+\sqrt{1-\tau}}{1-\sqrt{1-\tau}}\right) -
      i \pi\right]^2 & \tau < 1\end{matrix}\right. \;.
\eeq
The coupling $\alpha_s$ is evaluated at the scale $M_{H_i}$. The sums
are taken over the top, bottom and charm quarks and over all squark
mass eigenstates. \s

\noindent \underline{Decays into photons and into $Z\gamma$:} 
The decays into photons are
mediated by charged particle loops. In addition to the Feynman rules
introduced for the gluonic decay modes we have the Feynman rules for
the Higgs couplings to the charged gauge bosons $W^\pm$,
\beq
i g M_W g_{H_iWW} \;, \qquad \mbox{with } \qquad
g_{H_i WW} = {\cal R}_{i1} \cos \beta + {\cal R}_{i2} \sin \beta \;, 
\eeq
and for the neutral Higgs couplings to a charged Higgs bosons pair,
\beq
- \frac{ig M_Z^2}{2 M_W} g_{H_i H^+ H^-} \;.
\eeq
The coupling factor, which is
rather lengthy, can easily be read off from the program
code. Furthermore, we define the Higgs couplings to a chargino pair as
($j=1,2$)\footnote{Note that $g^P_{H_i \tilde{\chi}^+_j
    \tilde{\chi}^-_j}$ in the code is actually introduced as $i g^P_{H_i
    \tilde{\chi}^+_j \tilde{\chi}^-_j}$.}
\beq
-i \frac{g}{2} \left[ g^S_{H_i \tilde{\chi}^+_j
    \tilde{\chi}^-_j} -i \gamma_5 \, g^P_{H_i \tilde{\chi}^+_j
    \tilde{\chi}^-_j} \right] \;.
\eeq
Only the diagonal couplings appear in the partial
width. Their explicit form can be inferred from the program code. The
leading order decay width into photons is then given by
\begin{align}
  \Gamma(H_i \to \gamma\gamma) = \frac{G_F \alpha^2
    M_{H_i}^3}{128\pi^3 \sqrt{2}} & \left(\, \left| \sum_{f} N_{cf} e_f^2\,
      g^S_{H_i ff}  A^S_{1/2}(\tau_f) + \sum_{\tilde{\chi}^\pm_j}
      \frac{M_W}{M_{\tilde{\chi}_j^\pm}} g^S_{H_i \tilde{\chi}^+_j
    \tilde{\chi}^-_j} A^S_{1/2} (\tau_{\tilde{\chi}^\pm_j}) 
    \right. \right. \nonumber\\
    &  \left. \left. \hspace*{-0.7cm} + g_{H_i WW} A_1
      (\tau_W) + \frac{M_Z^2}{2 M_{H^\pm}^2} g_{H_i H^+
          H^-} A_0 (\tau_{H^\pm}) +  
    \sum_{\tilde{f}} N_{cf} e_f^2 \frac{M_Z^2}{m^2_{\tilde{f}}}\,  g_{H_i
          \tilde{f} \tilde{f}} A_0 (\tau_{\tilde{f}}) \right|^2 
        \nonumber \right. \\
    & \left.  \hspace*{-0.7cm} + 4
      \left| \sum_{f} N_{cf} e_f^2 \, g^P_{H_i ff}  A^P_{1/2}(\tau_f) 
    + \sum_{\tilde{\chi}^\pm_j} \frac{M_W}{M_{\tilde{\chi}_j^\pm}} 
     g^P_{H_i \tilde{\chi}^+_j
    \tilde{\chi}^-_j} A^P_{1/2} (\tau_{\tilde{\chi}^\pm_j}) \right|^2\, \right) \;.
\end{align}
The fermion sum proceeds over the top, charm and bottom quarks as well as
the $\tau$ lepton. The sum over $\tilde{f}$ includes all squark and
the stau mass eigenstates. With $N_{cf} =3, 1$ we denote the
colour factor of the (s)quarks and the (s)tau, respectively, and with
$e_f$ their corresponding electric charge. The loop function $A_1$ is
given by \cite{vainshteineal}
\beq
A_1 (\tau)= - \left( 2 + 3\tau + 3\tau (2-\tau) f(\tau)
  \right) \;.
\eeq
Analogously, the decay width into $Z\gamma$ can be derived from the
decay widths of the scalar and pseudoscalar Higgs bosons in the real
NMSSM with the appropriate replacements by the couplings as given in
this subsection. \s

\noindent \underline{Other decay modes:} In the decay widths
into bosonic final states (gauge bosons, Higgs pairs, Higgs
plus gauge boson pairs, sfermions) the respective coupling factors simply
have to be replaced with the corresponding couplings of the
CP-violating Higgs bosons. In case of complex valued couplings the
absolute value squared has to be taken in the decay width. The decay widths
into the fermionic final states (quark and charged fermion pairs as
well as chargino and neutralino final states) are given by the
incoherent sum of the scalar and pseudoscalar decay widths of
the real NMSSM. In the decays into bottom
and strange quarks as well as into tau leptons higher order SUSY
corrections have been included through effective Yukawa couplings. The
formulae are given in section \ref{sec:deltab}. \s

\noindent \underline{Charged Higgs boson decays:} The decays into
quark and lepton pairs do not change with respect to the real
NMSSM. The decay widths into a charged $W$ and neutral Higgs $H_i$ final
state are derived from the CP-conserving case by adding up
incoherently the decay widths into $W$ plus CP-even and $W$ plus CP-odd
Higgs bosons, respectively. The decay widths into sfermions are the
same as in the real NMSSM but with the respective coupling squared
replaced by the absolute value squared of the now complex valued
coupling. Also the charged Higgs couplings to a neutralino-chargino
pair are in general complex. Defining the Feynman rule as
\beq
- \frac{ig}{2} \left[ a_L P_L + a_R P_R \right] \;,
\eeq 
with the projectors $P_{L,R}= (1\mp \gamma_5)/2$ and complex $a_L,
a_R$ ({\it cf.} the program code for the explicit 
expressions), the decay width is given by
\beq
\Gamma (H^\pm \to \tilde{\chi}^\pm_j \tilde{\chi}^0_k) = \frac{G_F
  M_W^2 \, \lambda (M_{\tilde{\chi}^\pm_j}^2,M_{\tilde{\chi}^0_k}^2;
    M_{H^\pm}^2)}{2\sqrt{2} \pi M_{H^\pm}} && \hspace*{-0.6cm}  
\left[ \left(|a_L|^2 +
  |a_R|^2\right) \left(M_{H^\pm}^2 - M_{\tilde{\chi}^\pm_j}^2 -
  M_{\tilde{\chi}^0_k}^2 \right) \right. \nonumber\\ 
&& \hspace*{-0.6cm} \left. - 4 \, \Re (a_L^* a_R)  \,M_{\tilde{\chi}^\pm_j} 
  M_{\tilde{\chi}^0_k}\right] \;,
\eeq
with $j=1,2$ and $k=1,...,5$ and the two-body phase space function
$\lambda(x,y;z)$ given in Eq.~(\ref{eq:2body}).

\subsection{SUSY corrections to decays into
  fermions in the real and in the complex NMSSM \label{sec:deltab}}
The leading parts of the SUSY--QCD and SUSY--elw corrections to
the decays into a bottom quark pair can be taken into account by
absorbing them into effective bottom Yukawa couplings. The leading
corrections can be obtained from an effective Lagrangian
\cite{Carena:1999py,Guasch:2003cv} and can be taken over from the
MSSM case by deriving the effective 
Lagrangian for the NMSSM Higgs sector. In the real NMSSM it is given by
\beq
{\cal L}_{eff} = - y_b \overline{{b}}_R \left[ (1+\Delta_1) H_d^0 +
  \frac{\lambda (1+\Delta_1) \Delta_b}{\mu_{\scriptsize
      \mathrm{eff}}\tan\beta} S^* H_u^{0*} \right] b_L + h.c. \;, 
\eeq
where $H_{d(u)}^0$ are the neutral components of the doublet fields
coupling to down-type (up-type) quarks. The corrections $\Delta_b$ and
$\Delta_1$ include the SUSY--QCD and SUSY--elw corrections and   
induce a modification of the relation between the 
bottom quark mass $m_b$ and the Yukawa coupling $y_b$. 
After expanding the Lagrangian to include the higher order corrections
we have
\beq
{\cal L}_{eff} &=& - m_b \bar{b} \left[ 1 -i \gamma_5 \frac{G}{v}
\right] b \nonumber \\
&& - \frac{m_b/v}{1+\Delta_b} \bar{b} \left\{ \left[ \frac{{\cal
      R}^S_{11}}{\cos\beta} \left( 1 + \Delta_b \, \frac{{\cal
        R}^S_{12}}{{\cal R}^S_{11}} \frac{1}{\tan\beta} \right) +
  \left( \frac{{\cal R}^S_{13} v}{v_s} \right) \Delta_b  \right] H_1
\right. \nonumber \\
&& \phantom{- \frac{m_b/v}{1+\Delta_b} \bar{b}} +
\left[ \frac{{\cal
      R}^S_{21}}{\cos\beta} \left( 1 + \Delta_b \, \frac{{\cal
        R}^S_{22}}{{\cal R}^S_{21}} \frac{1}{\tan\beta} \right) +
  \left( \frac{{\cal R}^S_{23} v}{v_s} \right) \Delta_b  \right] H_2
\nonumber \\
&& \phantom{- \frac{m_b/v}{1+\Delta_b} \bar{b}} +
\left[ \frac{{\cal
      R}^S_{31}}{\cos\beta} \left( 1 + \Delta_b \, \frac{{\cal
        R}^S_{32}}{{\cal R}^S_{31}} \frac{1}{\tan\beta} \right) +
  \left( \frac{{\cal R}^S_{33} v}{v_s} \right) \Delta_b  \right] H_3
\nonumber \\
&& \phantom{- \frac{m_b/v}{1+\Delta_b} \bar{b}} -
\left[ {\cal R}^P_{11} \tan\beta \left( 1 - \Delta_b \,
    \frac{1}{\tan^2\beta} \right) - 
  \left( \frac{{\cal R}^P_{12} v}{v_s} \right) \Delta_b  \right] i
\gamma_5 A_1 \nonumber \\
&& \phantom{- \frac{m_b/v}{1+\Delta_b} \bar{b}} \left. -
\left[ {\cal R}^P_{21} \tan\beta \left( 1 - \Delta_b \,
    \frac{1}{\tan^2\beta} \right) - 
  \left( \frac{{\cal R}^P_{22} v}{v_s} \right) \Delta_b  \right] i
\gamma_5 A_2 \right\} b\; ,
\eeq
where we have used 
\beq
m_b &=& \frac{y_b v_d}{\sqrt{2}} (1+\Delta_b)(1+\Delta_1) \\
H_d^0 &=& \frac{1}{\sqrt{2}} [v_d + {\cal R}^S_{11} H_1 + {\cal
  R}^S_{21} H_2 + {\cal R}^S_{31} H_3 + i {\cal R}^P_{11} \sin\beta
A_1 + i {\cal R}^P_{21} \sin\beta A_2 + i G \cos \beta] \\
H_u^0 &=& \frac{1}{\sqrt{2}} [v_u + {\cal R}^S_{12} H_1 + {\cal
  R}^S_{22} H_2 + {\cal R}^S_{32} H_3 + i {\cal R}^P_{11} \cos\beta
A_1 + i {\cal R}^P_{21} \cos\beta A_2 - i G \sin \beta] \\
S &=& \frac{1}{\sqrt{2}} [v_s + {\cal R}_{13}^S H_1 + {\cal R}_{23}^S
H_2 + {\cal R}_{33}^S H_3 + i {\cal R}_{12}^P A_1 + i {\cal R}_{22}^P A_2]
\; ,
\eeq
with the mixing matrices ${\cal R}^S$ and ${\cal R}^P$
defined in Eq.~(\ref{eq:Srotation}) and (\ref{eq:Protation}),
respectively. The correction $\Delta_b$ contains the one-loop
SUSY--QCD and SUSY--elw corrections,
\beq
\Delta_b &=& \frac{\Delta_b^{QCD}+\Delta_b^{elw}}{1+\Delta_1} 
\label{eq:deltabcorr} \\
\Delta_b^{QCD} &=& \Delta_b^{QCD(1)} \\
\Delta_b^{elw} &=& \Delta_b^{elw(1)} \;,
\eeq
with the one-loop corrections given by 
\beq
\Delta_b^{QCD (1)} &=& 
\frac{C_F}{2} \, \frac{\alpha_s (\mu_R)}{\pi} \, m_{\tilde{g}} \,
\mu_{\mbox{\scriptsize eff}} \, \tan\beta \, I (m_{\tilde{b}_1}^2,
m_{\tilde{b}_2}^2 , m_{\tilde{g}}^2 )  \label{eq:deltaqcd1} \\
\Delta_b^{elw (1)} &=& \frac{y_t^2 (\mu_R)}{(4\pi)^2} \, A_t \,
\mu_{\mbox{\scriptsize eff}} \, \tan\beta \, I (m_{\tilde{t}_1}^2, 
m_{\tilde{t}_2}^2 , \mu_{\mbox{\scriptsize eff}}^2) \;,
\eeq
where $y_t = \sqrt{2} m_t /(v \sin\beta)$ denotes the top-Yukawa
coupling, $C_F=4/3$, and 
\beq
\Delta_1 = - \frac{C_F}{2} \frac{\alpha_s (\mu_R)}{\pi}\, m_{\tilde{g}} \,
A_b \, I (m_{\tilde{b}_1}^2,
m_{\tilde{b}_2}^2 , m_{\tilde{g}}^2 ) \; . \label{eq:delta1}
\eeq 
The generic function $I$ is defined as
\beq
I(a,b,c) &=& \frac{ab \, \log \frac{a}{b} + bc \, \log \frac{b}{c} +
  ca \, \log \frac{c}{a}}{(a-b)(b-c)(a-c)} \;.
\eeq
Note that the scale of $\alpha_s$ in the SUSY--QCD corrections
has been set equal to
$\mu_R=(m_{\tilde{b}_1}+m_{\tilde{b}_2}+|M_{\tilde{g}}|)/3$, while in
the SUSY--elw corrections it is
$\mu_R=(m_{\tilde{t}_1}+m_{\tilde{t}_2}+|\mu|)/3$, each
$\alpha_s$ calculated with five active flavours. The higher order
corrected decay width of the NMSSM Higgs bosons
$\Phi=H_1,H_2,H_3,A_1,A_2$ into $b\bar{b}$, including QCD and SUSY--QCD
corrections, can be cast into the form \cite{Guasch:2003cv}
\beq
\Gamma (\Phi \to b\bar{b}) = \frac{3 G_F M_\Phi}{4 \sqrt{2} \pi} \,
\overline{m}_b^2 (M_\Phi) \, [\Delta_{\mbox{\scriptsize QCD}} +
\Delta_t^\Phi] \, \tilde{g}_b^\Phi \, [\tilde{g}_b^\Phi +
\Delta^{rem}_{SQCD}] \;. \label{eq:correctedgambb} 
\eeq 
The logarithmically enhanced
part of the QCD corrections has been absorbed in the running
$\overline{\mbox{MS}}$ bottom quark mass $\overline{m}_b (M_\Phi)$ at the
corresponding Higgs mass scale $M_\Phi$. The QCD corrections
$\Delta_{QCD}$ and the top quark induced corrections $\Delta_t^\Phi$
can be taken over from the MSSM case
\cite{nearthreshold,abovethreshold,chetyrkin} by adapting the Higgs
couplings and read\footnote{Note that actually in the code we have
  programmed the QCD corrections for the completely massive case at
  next-to-leading order, translated to the $\overline{\mbox{MS}}$
  scheme, and interpolated with the massless expression for large Higgs
masses, according to the implementation in {\tt HDECAY} \cite{hdecay}.}
\begin{eqnarray}
\Delta_{\rm QCD} & = & 1 + 5.67 \frac{\alpha_s (M_\Phi)}{\pi} + (35.94 -
1.36
N_F) \left( \frac{\alpha_s (M_\Phi)}{\pi} \right)^2 \nonumber \\
& & + (164.14 - 25.77 N_F + 0.259 N_F^2) \left(
\frac{\alpha_s(M_\Phi)}{\pi}
\right)^3 \nonumber \\
\Delta_t^{\phi_S} & =&
\frac{g_t^{\phi_S}}{g_b^{\phi_S}}~\left(\frac{\alpha_s (M_{\phi_S})}{\pi}
\right)^2 \left[ 1.57 - \frac{2}{3} \log \frac{M_{\phi_S}^2}{M_t^2}
+ \frac{1}{9} \log^2 \frac{\overline{m}_b^2
(M_{\phi_S})}{M_{\phi_S}^2}\right]\nonumber \\
\Delta_t^{\phi_P} & = &
\frac{g_t^{\phi_P}}{g_b^{\phi_P}}~\left(\frac{\alpha_s (M_{\phi_P})}{\pi} 
\right)^2
\left[ 3.83 - \log \frac{M_{\phi_P}^2}{M_t^2} + \frac{1}{6} \log^2
\frac{\overline{m}_b^2 (M_{\phi_P})}{M_{\phi_P}^2} \right] \;,
\end{eqnarray}
with $\phi_S \equiv H_1,H_2,H_3$, $\phi_P \equiv A_1,A_2$ and where
$N_F=5$ active flavors are taken into account. The coupling factors
$g^\Phi_b$ ($g^\Phi_t$) with respect to the SM Higgs-bottom (top) Yukawa coupling are given by
\beq
g^{H_i}_t &=& {\cal R}^S_{i2} / \sin\beta \;, \qquad g^{H_i}_b = {\cal
  R}^S_{i1} / \cos\beta \; , \qquad i=1,2,3 \;, \label{eq:hbcoup}\\
g^{A_j}_t &=& {\cal R}^P_{j1}/\tan\beta \;, \qquad 
g^{A_j}_b = {\cal R}^P_{j1} \tan\beta \; , \qquad j=1,2 \;.
\label{eq:abcoup}
\eeq
The dominant part of the SUSY--QCD
\cite{Guasch:2003cv,coarasa,Eberl:1999he} 
corrections has been absorbed in the effective Yukawa couplings
$\tilde{g}^\Phi_b$. Adapting the results from the MSSM to the NMSSM they
read
\beq
\tilde{g}_b^{H_1} &=& \frac{g^{H_1}_b}{1 + \Delta_b} \left[ 1 +
  \Delta_b \left( \frac{{\cal R}^S_{12}}{{\cal R}^S_{11} \tan\beta} +
    \frac{{\cal R}^S_{13} v\cos\beta}{{\cal R}^S_{11} v_s} \right)
\right] \label{eq:effcoup1} \\
\tilde{g}_b^{H_2} &=& \frac{g^{H_2}_b}{1 + \Delta_b} \left[ 1 +
  \Delta_b \left( \frac{{\cal R}^S_{22}}{{\cal R}^S_{21} \tan\beta} +
    \frac{{\cal R}^S_{23} v\cos\beta}{{\cal R}^S_{21} v_s} \right)
\right] \\
\tilde{g}_b^{H_3} &=& \frac{g^{H_3}_b}{1 + \Delta_b} \left[ 1 +
  \Delta_b \left( \frac{{\cal R}^S_{32}}{{\cal R}^S_{31} \tan\beta} +
    \frac{{\cal R}^S_{33} v\cos\beta}{{\cal R}^S_{31} v_s} \right)
\right] 
\eeq
and
\beq
\tilde{g}_b^{A_1} &=& \frac{g^{A_1}_b}{1 + \Delta_b} \left[ 1 +
 \Delta_b \left( -\frac{1}{\tan^2\beta} -
   \frac{{\cal R}^P_{12} v}{{\cal R}^P_{11} v_S\tan\beta} \right) \right]
\\
\tilde{g}_b^{A_2} &=& \frac{g^{A_2}_b}{1 + \Delta_b} \left[ 1 +
 \Delta_b \left( -\frac{1}{\tan^2\beta} -
   \frac{{\cal R}^P_{22} v}{{\cal R}^P_{21} v_S\tan\beta} \right)
\right] \;.
\label{eq:effcoup5}
\eeq
The remaining part of the SUSY--QCD corrections, after the main
corrections have been absorbed in the effective bottom Yukawa
couplings $\tilde{g}_b^\Phi$, is given by the remainder $\Delta_{SQCD}^{rem}$.
The decay width of Eq.~(\ref{eq:correctedgambb}) has been
implemented in the decay program. \s

The SUSY--QCD corrections at one-loop order have also been included in
the decays into strange quarks, {\it i.e.}~$\Delta_b$ of Eq.~(\ref{eq:deltabcorr})
for these decays is replaced by
\beq
\Delta_s = \left. \frac{\Delta_b^{QCD(1)}}{1+\Delta_1}\right|_{b\to s}
\;, \label{eq:deltabstrange} 
\eeq
with $\Delta_b^{QCD(1)}$ and $\Delta_1$ obtained from
Eqs.~(\ref{eq:deltaqcd1}) and (\ref{eq:delta1}) after substituting 
$A_b, m_{\tilde{b}_{1,2}}$ with $A_s, m_{\tilde{s}_{1,2}}$.
For the decays into lepton finals states $l=e,\mu,\tau$ the SUSY
corrections have been included in the decay widths by absorbing them
into the effective couplings $\tilde{g}_l^\Phi$,
\beq
\Gamma (\Phi \to ll) = \frac{G_F}{4 \pi \sqrt{2}} \, M_\Phi \, 
  m_l^2 \,(\tilde{g}_l^{\Phi})^2 \;.
\eeq
The effective couplings $\tilde{g}_l^\Phi$ are defined as in
Eqs.~(\ref{eq:effcoup1})-(\ref{eq:effcoup5}) by replacing $b$ with $l$
and with $\Delta_l$ given by \cite{Pierce:1996zz}
\beq
\Delta_l = \frac{\alpha_1}{4\pi} M_1 \mu_{\scriptsize \mbox{eff}}
\tan\beta \, I (m_{\tilde{l}_1}^2,m_{\tilde{l}_2}^2,M_1^2) +
\frac{\alpha_2}{4 \pi} M_2 \mu_{\scriptsize \mbox{eff}} \tan\beta \, 
I(m_{\tilde{\nu}_l}^2,M_2^2, \mu_{\scriptsize \mbox{eff}}^2) 
\label{eq:deltalexp}
\eeq
with $\alpha_{1,2} = g_{1,2}^2/4\pi$. \s

The SUSY--QCD and SUSY--elw corrections to the charged Higgs boson decays
$H^\pm$ into up bottom, charm bottom and top bottom final states,
respectively, have been included 
analogously to the decays of the neutral Higgs bosons $\Phi$ into
$b\bar{b}$, Eq.~(\ref{eq:correctedgambb}), with $\Delta_{\scriptsize
  \mbox{QCD}}$ given in \cite{Djouadi:1994gf} and $\Delta_t^\Phi=0$.
The effective coupling $\tilde{g}^{H^\pm}_b$ reads
\beq
\tilde{g}^{H^\pm}_b = \frac{g^{H^\pm}_b}{1+\Delta_b} \left[ 1 -
  \frac{\Delta_b}{\tan^2\beta} \right] \;,
\eeq
with
\beq
g^{H^\pm}_b = \tan\beta
\eeq
and $\Delta_b$ given in Eq.~(\ref{eq:deltabcorr}). The SUSY--QCD
corrections have been taken into account in the decays of $H^\pm$ into
a strange quark with an up, charm and top quark, respectively, by
replacing $\Delta_b$ of 
Eq.~(\ref{eq:deltabcorr}) with $\Delta_s$ of Eq.~(\ref{eq:deltabstrange}).
Finally, the SUSY corrections are implemented in the $H^\pm$
decays into $l \nu_l$ ($l=e,\mu,\tau$) via the effective couplings
given in terms of $\Delta_l$. \s

In the complex NMSSM the leading parts of the
SUSY--QCD and SUSY--elw corrections in the decays into bottom
quarks are derived from the effective Lagrangian
\beq
{\cal L}_{eff} = - y_b  \overline{{b}}_R\left[(1+\Delta_{1})
 H_d^0 + \frac{\lambda^* e^{i \varphi_u} (1+\Delta_1)
   \Delta_b}{\mu_{\scriptsize  
\mathrm{eff}}^*\tan\beta} S^* H_u^{0*} \right] b_L   + h.c. \;.
\label{eq:eff_lagrangian_cp}
\eeq
Introducing the neutral components of the two doublets and singlet
fields as
\bea
 H_d^0 &=&\fr{1}{\sqrt{2}}\left[v_d+\sum_{j=1}^{5}( {\cal R}_{j1}+ 
i \sin\beta \,  {\cal R}_{j4}) H_j  + i \cos\beta G \right], \\
 H_u^0 &=&\fr{e^{i\varphi_u}}{\sqrt{2}}\left[v_u+\sum_{j=1}^{5}( {\cal R}_{j2}+
 i \cos\beta \, {\cal R}_{j4}) H_j  - i \sin\beta G \right], \\
 S &=&\fr{e^{i\varphi_s}}{\sqrt{2}}\left[v_s+\sum_{j=1}^{5}( {\cal R}_{j3}
 + i {\cal R}_{j5}) H_j  \right], 
\eea
in Eq.~(\ref{eq:eff_lagrangian_cp}), the Lagrangian reads
\beq
{\cal L}_{eff} &=& - m_b \bar{b} \left( 1 -i \gamma_5 \frac{G}{v}
\right) b  -\frac{m_b}{v} \sum_{j=1}^{5} \bar{b} \left[  \tilde{g}_{bL}^{H_{j}} P_{L} + (\tilde{g}_{bL}^{H_{j}})^* P_R) \right] H_j \, b,
\eeq
with 
\beq
 \tilde{g}_{bL}^{H_{j}}= \fr{1}{(1+ \Delta_b)} \left[ \fr{{\cal R}_{j1}}{\cos\beta} 
+  \fr{{\cal R}_{j2}}{\sin\beta} \Delta_b
+  \fr{ {\cal R}_{j3}v}{v_s}\Delta_b + i {\cal R}_{j4}\tan\beta\left( 1- \fr{\Delta_b}{\tan^2\beta}\right)
-i  \fr{ {\cal R}_{j5}v}{v_s}\Delta_b\right] \;.
\eeq
The rotation matrix ${\cal R}$ has been defined in
Eq.~(\ref{eq:cpR}). 
The correction
$\Delta_b$ includes the one-loop SUSY--QCD and SUSY--elw corrections
which in the complex NMSSM are given by
\beq
\Delta_b &=& \frac{\Delta_b^{QCD(1)}  + \Delta_b^{elw(1)}
}{1+\Delta_1}\;,  \label{eq:deltabcomplex} \\
\Delta_b^{QCD (1)} &=& 
\frac{C_F}{2} \, \frac{\alpha_s (\mu_R)}{\pi} \, M_3^* \,
\mu_{\mbox{\scriptsize eff}}^* \, \tan\beta \, I (m_{\tilde{b}_1}^2,
m_{\tilde{b}_2}^2 , m_{\tilde{g}}^2 ) \;, \label{eq:deltabqcd1complex}
\\ 
\Delta_b^{elw (1)} &=& \frac{y_t^2 (\mu_R)}{(4\pi)^2} \, A_t^* \,
\mu_{\mbox{\scriptsize eff}}^* \, \tan\beta \, I (m_{\tilde{t}_1}^2, 
m_{\tilde{t}_2}^2 , |\mu_{\mbox{\scriptsize eff}}|^2) \;,\\
\Delta_1&=& - \frac{C_F}{2} \frac{\alpha_s}{\pi} \, M_3^*\,
A_b \, I (m_{\tilde{b}_1}^2,
m_{\tilde{b}_2}^2 , m_{\tilde{g}}^2 )\;. 
\label{eq:deltabcorr_cp}
\eeq
The corrections $\Delta_b$ and $\Delta_1$ are in general complex and
depend on the phases of the gaugino mass $M_3$, of
the trilinear couplings $A_t$, $A_b$ and on the phase of the
effective Higgs mixing parameter, which in the CP-violating NMSSM is
given by $\mu_{\mbox{\scriptsize eff}}=\lambda v_s e^{i\varphi_s}/\sqrt{2}$. \s

The decay width of the Higgs boson $\Phi= H_{j}$ ($j=1,...,5$) into $b
\bar b$, in the complex NMSSM can be written as
\beq
\Gamma (\Phi \to b\bar{b}) = \frac{3 G_F M_\Phi}{4 \sqrt{2} \pi} \,
\overline{m}_b^2 (M_\Phi) \, [\Delta_{\mbox{\scriptsize QCD}} +
\Delta_t^\Phi] \, |\tilde{g}_{bL}^\Phi|^2 \;.
\label{eq:correctedgambb_cp} 
\eeq 
Note, in particular, that contrary to the real NMSSM we do not include
a remainder $\Delta^{rem}_{SQCD}$ in the decay width, as it is not
available at present. This leads, when using the program
package for the complex NMSSM in the limit of the real NMSSM to
differences in the decay 
widths into $b$-quark pairs with respect to the results obtained from
the program package for the real NMSSM. The differences are below the percent
level.\footnote{Further differences appear in the decays
  into stop and sbottom pairs, respectively, as in the complex case we
do not include SUSY--QCD corrections.}
The one-loop SUSY--QCD corrections to the decays into strange
quarks are obtained after substituting $\Delta_b$ as given in
Eq.~(\ref{eq:deltabcomplex}) with $\Delta_s =
\Delta_b^{QCD(1)}/(1+\Delta_1)|_{b\to s}$. For the
decays into leptons $\Delta_b$ has to be replaced with $\Delta_l$,
where $\Delta_l$ in the complex case is given by Eq.~(\ref{eq:deltalexp})
with $M_1,M_2$ and $\mu_{\text{eff}}$ replaced by $M_1^*,M_2^*$ and
$\mu_{\text{eff}}^*$ and $M_1^2$, $M_2^2$, $\mu_{\text{eff}}^2$ by
$|M_1|^2$, $|M_2|^2$, $|\mu_{\text{eff}}|^2$. \s

The effective coupling (including the $\Delta_b$ effect) of the charged Higgs 
boson to a top-bottom quark pair can be read off from the
Lagrangian
\beq
 {\cal L}_{eff} = \fr{\sqrt 2}{v}(V^{\text{CKM}}_{tb})^*\bar b\left[
   m_{b}\tan\beta 
\frac{1- \Delta_b/\tan^2\beta}{1+\Delta_b} P_L + \frac{m_{t}}{\tan\beta}
P_R \right]H^- t+ h.c. \;,
\eeq
where $V^{\rm CKM}$ denotes the CKM matrix and the phase $\varphi_u$ 
has been included in the CKM matrix element through a factor $e^{i \varphi_u/2}$.

\section{Program Description \label{sec:progdescr}} 
The program package consists of a wrap file called {\tt
  nmssmcalc.f} and three main files: 
\begin{enumerate}
\item {\tt CalcMasses.F} for the
calculation of the one-loop corrected NMSSM Higgs boson masses in the
real and the complex NMSSM; 
\item {\tt bhdecay.f} 
for the calculation of the NMSSM Higgs boson decay widths and
branching ratios in the real NMSSM; 
\item {\tt bhdecay\_c.f} for the
calculation of the decay widths and branching ratios in the complex
NMSSM.
\end{enumerate} 
There are additional files 
containing subroutines needed for the calculation of 
the decay widths. All files are written in Fortran. The package needs
two input files:
\begin{enumerate}
\item an input file in the SLHA format with default name {\tt
    inp.dat}, which is read in by {\tt nmssmcalc.f};
\item an input file {\tt bhdecay.in} which is read in by  {\tt
    bhdecay(\_c).f}  and which contains the setting of the CKM 
parameters as well as of several flags for the decay calculation. They
are specified in Appendix~\ref{sec:bmhdinp}.
\end{enumerate}
The mass routine {\tt CalcMasses.F} provides an output file in the SLHA
format, which is read in by the decay routine {\tt bhdecay(\_c).f}
which in turn writes out the results in an SLHA file. 
The whole package is compiled with the help of a {\tt makefile}. In
the following the various files and their functions will be described
as well as the compilation and the running of the program package. \s

\underline{{\tt nmssmcalc.f}:} This wrap file reads in the
input parameters needed for the calculation of the one-loop corrected
NMSSM Higgs boson masses. The input file which has the default name
{\tt inp.dat}, must be provided in the SLHA format and has to contain the
blocks {\tt MODSEL}, {\tt SMINPUTS}, {\tt MINPAR} and {\tt EXTPAR}, 
as well as {\tt IMEXTPAR} and {\tt CMPLX} in the complex 
case, with the related input parameters. The required parameters are
described in more detail in the following section
{\ref{sec:slha}}. The file then calls {\tt CalcMasses.F} by
transferring the parameters which have been read in, so that the routine can 
calculate the one-loop corrections of the NMSSM Higgs boson
masses and writes them out in an SLHA output file with default name {\tt
  slha.in}. Subsequently {\tt nmssmcalc.f} calls {\tt
  bhdecay.f}, in case the CP-conserving NMSSM has been 
chosen in the input file and {\tt bhdecay\_c.f} in case of a
complex NMSSM choice, respectively. These routines read in the output
file provided by {\tt CalcMasses.F} and calculate the Higgs decay widths and
branching ratios in the framework of the CP-conserving or CP-violating
NMSSM with the results written out in an SLHA output file (default name {\tt
  slha\_decay.out}). The user can also specify the names of the input
file and of the output files provided by the mass and decay routines
in the command line when running the program. \s

\underline{{\tt CalcMasses.F}:} This Fortran code contains the subroutine
{\tt CalcMasses} for the calculation
the one-loop corrections to the NMSSM Higgs boson masses in the
CP-conserving or CP-violating NMSSM. The renormalisation schemes
applied are a mixture of on-shell and $\overline{\mbox{DR}}$ conditions,
which are described in detail in Refs.~\cite{Ender:2011qh,Graf:2012hh} and
have been briefly sketched in sections~\ref{sec:cpconsloopcorrections}
and \ref{sec:cpviolloopcorrections}. As described in sections
\ref{sec:cpconsloopcorrections} 
and \ref{sec:cpviolloopcorrections} the schemes have been slightly modified in
order to match the requirements of the SLHA (see also
section~\ref{sec:slha}). The subroutine calls further subroutines
for the calculation of all NMSSM tree-level 
masses, of the one-loop Higgs masses, of the Higgs couplings, of the
counterterms etc. They are listed together with their 
functions in the header of {\tt CalcMasses.F}. At the end of the
calculation it provides an SLHA output file {\tt slha.in}. This file
contains in particular the loop-corrected Higgs boson masses with the
related Higgs mixing parameters, the NMSSM SUSY particle spectrum with
corresponding mixing matrices as well as further blocks needed
for the calculation of the decay widths.  \s
%

\underline{{\tt bhdecay.f, bhdecay\_c.f}:} The subroutines contained
in these Fortran codes 
calculate the decay widths and branching ratios of the neutral
and charged NMSSM Higgs bosons in the CP-conserving and the CP-violating
case, respectively. They closely follow the structure and approximations
 of the Fortran code {\tt
  HDECAY} version 6.00 
\cite{hdecay}\footnote{See \cite{Contino:2013kra} for the extension of
  {\tt HDECAY} to effective Lagrangians for a light Higgs-like scalar.}, which
calculates the Higgs decay widths and branching 
ratios in the SM and the MSSM including the most important higher
order corrections. Thus the higher order corrections have been taken
over wherever they can be adapted to the NMSSM case. This also means
that a lot of common blocks and routines inherent in {\tt HDECAY} have been
retained. They are dubbed by the suffix {\tt \_HDEC}. Routines and
common blocks that have been extended to the NMSSM case or are
specific to the NMSSM case are denoted by the suffix {\tt \_NMSSM},
respectively {\tt \_CNMSSM} in the CP violating NMSSM. These Fortran
routines are supplemented by various Fortran files taken over from {\tt
  HDECAY} and adapted to the NMSSM where necessary: {\tt dmb.f}
(contains the 2-loop SUSY--QCD and
SUSY--elw corrections to the bottom Yukawa coupling in the real
MSSM\footnote{These 2-loop corrections have not been adapted to the NMSSM
  yet, and therefore are not included in the NMSSM decays at
  present. This will be done in a future updated version of the
  program so that we include this routine already now in the program
  package.}), {\tt hgaga.f} (needed for the QCD corrections to the
decays into photons), {\tt slha\_nmssm.f} (for reading in the
SLHA input file) and in case of the real NMSSM also {\tt hsqsq\_nmssm.f}
(called for the SUSY--QCD corrections to decays into squark
pairs).  Finally {\tt Xvegas.f} has been linked, which is used in
the numerical integration of some off-shell decays. \s

The general structure of {\tt bhdecay(\_c).f} is the following: After
calling the subroutine \\ {\tt read\_(c)nmssm.f} to read in the input
files {\tt bhdecay.in} and {\tt slha.in} it calls the core
routine for the decays, {\tt hdec\_(c)nmssm}, in which the decay
widths are calculated. This routine calls several other help
routines. Afterwards {\tt write\_(c)nmssm} is called to write out the
results in an SLHA output file. \s

\underline{How to compile and run the program package:} The program
package is compiled with the help of a {\tt makefile} by typing {\it
  make}. This provides an executable file called {\it
  run}. The user has the choice to provide the names of the input and
output files for {\tt CalcMasses.F} (first and second argument) and
the name of the output file provided by the decay routine (third
argument) in the command line. Hence the command will be {\it
  run name\_file1 name\_file2 name\_file3} in this case. \s

\underline{Webpage of the program package:} The program package can be
downloaded from the url:
\begin{center}
{\tt http://www.itp.kit.edu/$\sim$maggie/NMSSMCALC/}
\end{center}
This webpage also contains a short description of the
program. Furthermore, it informs about changes and updates of the 
program package. 

\section{The SUSY Les Houches Accord \label{sec:slha}}
The SUSY Les Houches Accord \cite{slha1} has been extended in
SLHA2 \cite{slha2} to include the case of the NMSSM as well as possible
CP-violation in the MSSM. In the following we list issues of our
program package related to SLHA2 and how we treated them. Furthermore
the parameters are specified, that have to be given by the user for
the calculation of the loop-corrected mass values and the calculation
of the branching ratios in the CP-conserving and CP-violating NMSSM. \s

The $2 \times 3$ mixing matrix $P$ for the rotation into the
pseudoscalar mass eigenstates defined in the SLHA2 for the 
CP-conserving NMSSM differs from
the matrix ${\cal R}^P$ rotating the imaginary parts of the
interaction states to the CP-odd mass eigenstates, {\it
  cf.}~Eq.~(\ref{eq:Protation}). The tree-level matrix elements are related
through
\beq
{\cal R}^P_{11} &=& P_{11}/\sin\beta_n = P_{12}/\cos \beta_n \qquad
{\cal R}^P_{12} = P_{13} \\
{\cal R}^P_{21} &=& P_{21}/\sin\beta_n = P_{22}/\cos \beta_n \qquad
{\cal R}^P_{22} = P_{23} \;.
\eeq
The angle $\beta_n$ is the rotation angle of the matrix ${\cal R}^G$ 
separating the massless Goldstone boson. It coincides at tree-level
with the angle $\beta$ defined through the ratio of the VEVs $v_u$ 
and $v_d$, $\tan\beta = v_u/v_d$. \s

In the block {\tt MODSEL} the user has to specify the model to be
used and if CP-violation is to be included or not, \s

\noindent\texttt{
\hspace*{0.2cm} BLOCK MODSEL \\
\hspace*{0.4cm}   3 1  \# NMSSM \\
\hspace*{0.4cm}   5 0  \# (0: CP conservation; 2: general CP violation)
} \s

\noindent 
Furthermore, the block {\tt SMINPUTS} has to be provided as well as,
in the block {\tt MINPAR}, the value for $\tan\beta$. Note that in the
input file {\tt inp.dat}, which is read in by {\tt nmssmcalc.f}, the 
block {\tt SMINPUTS} has been extended by the $W$ boson pole mass as
this mass value is used in the calculation of the loop-corrected Higgs
boson masses and of the decay widths. It has to be given in the 9th entry of
the block. \s

If working in the CP-conserving case it is sufficient to supply
the block {\tt EXTPAR}. The CP-violating case also requires the block
{\tt IMEXTPAR} which provides the imaginary parts to the corresponding
real parts.
In these two blocks the
gaugino soft SUSY breaking mass parameters $M_1$, $M_2$ and $M_3$ have
to be given and the soft SUSY breaking mass parameters for the 3rd
generation, $m_{Q_3}, m_{t_R}, m_{b_R}, m_{L_3}$ and $m_{\tau_R}$\footnote{These are always real.}. Also the
soft SUSY breaking trilinear couplings $A_t$, $A_b$, $A_\tau$,
$A_\lambda$ and $A_\kappa$, the
NMSSM couplings $\lambda$ and $\kappa$ and the effective $\mu$
parameter, $\mu_{\mbox{\scriptsize eff}}$, have to be set. 
Together with $\lambda$ the latter value determines the value of $v_s$,
and in the complex case also of $\varphi_s$, according to 
$\mu_{\mbox{\scriptsize eff}} = \lambda v_s e^{i \varphi_s}/\sqrt{2}
$. 
The values of the soft SUSY breaking trilinear couplings
of the second generation can be specified as well. If they are not given, they
are set equal to the corresponding values of the third generation. The
same holds for the first generation values: unless given they are set
equal to the corresponding values of the second generation. The
soft SUSY breaking masses of the first two generations are set
equal to the corresponding values of the third generation, if not
included in the SLHA input. \s

In our calculation of the loop-corrected masses we retain the option 
to either use $M_{H^\pm}$ or $\Re A_\lambda$. Depending on which input 
value is supplied by the user the according renormalisation scheme,
as described in sections \ref{sec:oneloopmass} and \ref{sec:complexoneloop}, 
is chosen for the calculation of the loop corrected masses. 
If the mass of the charged Higgs boson is chosen as an on-shell input
the real part of the corresponding $A_\lambda$
value in the $\overline{\mbox{DR}}$ scheme is calculated, which is
needed in the SLHA output file in accordance with the
SLHA conventions. Furthermore the $\overline{\mbox{DR}}$ values
of the imaginary parts of $A_\lambda$ and $A_\kappa$ are also 
supplied in the output file for the CP-violating case. \s

In the block {\tt EXTPAR} the user has the option to give the
renormalisation scale at which the parameters are evaluated. If this
value is not given the renormalisation scale $ \mu_{\mbox{\scriptsize
    ren}}$ is set equal to the
geometric mean of the soft SUSY breaking masses in the stop sector,
\beq
\mu_{\mbox{\scriptsize ren}} = \sqrt{m_{Q_3} m_{t_R}} \;.
\eeq
At present this scale is only used to set the renormalisation scale in the mass calculation. \s

In SLHA2 there is no block foreseen for complex phases in the
Higgs sector. We therefore added a new block called {\tt
  CMPLX} where the value of the phase $\varphi_u$ can be specified
according to \s

\noindent\texttt{
\hspace*{0.2cm} BLOCK CMPLX \\
\hspace*{0.4cm}   3  \# phiu 
} 
\s\s

\noindent
We also had to add a new block containing the matrix elements
of the $5\times 5$ Higgs mixing matrix of the CP-violating Higgs
sector. The block has been called {\tt NHMIXC} and contains the 
one-loop corrected mixing matrix elements $\mbox{ZH}_{ij}$
in the basis $i=(H_1,H_2,H_3,H_4,H_5)$,
$j=(h_d,h_u,h_s,a,a_s)$\footnote{Note that $(h_d,h_u,h_s,a,a_s,G)^T =
  {\cal R}^G (h_d,h_u,h_s,a_d,a_u,a_s)^T$}. It
corresponds to the FORTRAN format 
\beq
\mbox{(1x,I2,1x,I2,3x,1P,E16.8,0P,3x,'\#',1x,A)}
\eeq 
so that the structure of the block is as follows, \s

\noindent\texttt{
\hspace*{0.2cm} BLOCK NHMIXC \\
\hspace*{0.4cm}   1 1  value11  \# ZH(h1,hd) \\
\hspace*{0.4cm}   1 2  value12  \# ZH(h1,hu) \\
\hspace*{0.8cm} \vdots
} 
\s \s

In the SLHA output of the loop corrected complex NMSSM Higgs boson masses,
there is some peculiarity. The pdg code for these particles is
increased with ascending mass values. This means in particular, that, 
irrespective of its amount of pseudoscalar admixture, the third
heaviest (heaviest) Higgs boson $H_3$ ($H_5$) has the pdg
code 36 (46) which in the CP-conserving case is reserved for the pseudoscalar
mass eigenstates. \s

There are several warnings that are issued by the code for the mass
calculation. In case one of the obligatory parameters is 
missing a warning is issued with the name of the missing
parameter. If the real NMSSM has been chosen in {\tt MODSEL} and
non-zero imaginary parts are nevertheless filled in {\tt IMEXTPAR},
the user is warned that these should be zero and the program is
terminated. A further warning is issued if a different model than the 
NMSSM is chosen in the block {\tt MODSEL}, as the program only calculates the
higher order corrections to the NMSSM Higgs boson masses. 
If the CP-conserving or violating case has not been specified, the
user is notified by the program to fill in the information. In case
the imaginary parts of 
$A_\lambda$ and $A_\kappa$ have been given, the user is warned, that
these values are ignored. They are fixed in our
  renormalisation prescription through the tadpole conditions, and the
  thus obtained values are given out in the SLHA output to ensure
consistency. 
If both the real part of $A_\lambda$ and $M_{H^\pm}$ are set in the input 
file, a warning is issued to use either of these values. Furthermore,
the $M_{H^\pm}$ value is assumed to be the default input and the value
of $A_\lambda$ is ignored in this case. 

\section{Summary and Outlook \label{sec:summary}}
We have presented the program package {\tt NMSSMCALC} for the
calculation of the loop-corrected NMSSM Higgs boson masses and decay
widths in the CP-conserving and the CP-violating NMSSM. The Higgs
boson masses are calculated at one-loop order in a mixed
renormalisation scheme of on-shell and $\overline{\mbox{DR}}$
conditions. The decays include the most up-to-date higher order corrections.  
The program will be continuously updated to include the
state-of-the art results for Higgs boson masses and decays. In
particular we plan to extend the SUSY--QCD corrections to the decays 
into stops and sbottoms to the complex NMSSM. Furthermore, the two-loop corrections
to the Higgs boson masses as well as the higher-order corrections to
the trilinear Higgs self-couplings will be included. The program 
thus serves as a tool for making precise predictions for the masses and
the decay widths of the NMSSM Higgs bosons. Therefore it allows to
reliably interpret the experimental results and distinguish between the Higgs
sectors of models beyond the SM. 

\subsubsection*{Acknowledgments} 
J.B., R.G., M.M., D.T.N. and K.W. are supported by the DFG SFB/TR9  ``Computational
Particle Physics''. R.G. acknowledges financial support by the
``Landesgraduiertenf\"orderung des Landes Baden-W\"urt\-tem\-berg''. 

\section*{Appendix}
\begin{appendix}

\section{The input file {\tt bhdecay.in \label{sec:bmhdinp}}}
In the following the various input values of the file {\tt
  bhdecay.in} are specified. We have \\

\noindent
\underline{VTB, VTS, VTD, VCB, VCS, VCD, VUB, VUS, VUD}: The CKM
matrix elements $V_{tb}$, $V_{ts}$, $V_{td}$, $V_{cb}$, $V_{cs}$,
$V_{cd}$, $V_{ub}$, $V_{us}$ and $V_{ud}$.\\

\noindent
\underline{NNLO (M)}: If $=0$ then the ${\cal O} (\alpha_s)$ formula
is used for the conversion of the quark pole into the
$\overline{\mbox{MS}}$ masses. If $=1$, the ${\cal O} (\alpha_s^2)$
formula is used.\\

\noindent
\underline{ON-SHELL}: If $=0$ then the off-shell decays into
$\bar{t} t^*$, $\bar{b}t^*$, $\bar{s} t^*$, $\bar{d} t^*$, 
into a Higgs and off-shell gauge boson and into two off-shell Higgs
bosons are calculated. If $=1$, they are not included. \\

\noindent
\underline{ON-SH-WZ}: If $=0$, the double off-shell pair decays into
$W^*W^*$, $Z^*Z^*$ are calculated. If $=-1$, the double off-shell
decays are included below threshold, but the on-shell decays above. If
$=1$, only the single off-shell decays into $W^*W$, $Z^*Z$ are
calculated below threshold, and the on-shell decays above.\\

\noindent 
\underline{OFF-SUSY}: If =0, the decays (and loops) into SUSY particles
are included. If =1, they are excluded.\\

\noindent
\underline{NF-GG}: Number of light flavours (3, 4 or 5) included in the
gluonic decays.\\ 

\end{appendix}



\newpage
\begin{thebibliography}{999} 

\bibitem{:2012gk}
G.~Aad {\it et al.}  [ATLAS Collaboration],
  Phys.\ Lett.\ B {\bf 716} (2012) 1
  [arXiv:1207.7214 [hep-ex]];
G.~Aad {\it et al.}  [ATLAS Collaboration], ATLAS-CONF-2012-162.

\bibitem{:2012gu}
S.~Chatrchyan {\it et al.}  [CMS Collaboration],
  Phys.\ Lett.\ B {\bf 716} (2012) 30
  [arXiv:1207.7235 [hep-ex]];
S.~Chatrchyan {\it et al.}  [CMS Collaboration], CMS-PAS-HIG-12-045.

\bibitem{Goldstone} J. Goldstone, A. Salam and S. Weinberg,
  Phys. Rev. {\bf 127} (1962) 965; S. Weinberg, Phys. Rev. Lett. {\bf
    19} (1967) 1264;  S.L. Glashow, S. Weinberg, Phys. Rev. Lett. {\bf
    20} (1968) 224; A. Salam, Proceedings Of The Nobel Symposium,
  Stockholm 1968, ed. N. Svartholm. 

\bibitem{Higgs} P.W. Higgs, Phys. Lett. {\bf 12} (1964) 132; Phys.\
  Rev.\ Lett.\  {\bf 13 } (1964)  508 and 
  Phys. Rev. {\bf 145} (1966)  1156; F. Englert and R. Brout,
  Phys. Rev. Lett. {\bf 13} (1964) 321;
  G.S. Guralnik, C.R. Hagen and
  T.W. Kibble, Phys. Rev. Lett. {\bf 13} (1964) 585.

\bibitem{genNMSSM1} P. Fayet, Nucl. Phys. B \textbf{90} (1975) 104; Phys. Lett.
B \textbf{64} (1976) 159; Phys. Lett. B \textbf{69} (1977) 489 and Phys. Lett. B
\textbf{84} (1979) 416; H.P. Nilles, M. Srednicki and D. Wyler, Phys. Lett. B
\textbf{120} (1983) 346; J.M. Frere, D.R. Jones and S. Raby, Nucl. Phys. B
\textbf{222} (1983) 11; J.P. Derendinger and C.A. Savoy, Nucl. Phys. B
\textbf{237} (1984) 307;  A.I. Veselov, M.I. Vysotsky and K.A. Ter-Martirosian,
Sov. Phys. JETP \textbf{63} (1986) 489; J.R. Ellis, J.F. Gunion, H.E. Haber, L.
Roszkowski and F. Zwirner, Phys. Rev. D \textbf{39}  (1989) 844; M. Drees, Int.
J. Mod. Phys. A \textbf{4}  (1989) 3635.

\bibitem{genNMSSM2} U. Ellwanger, M. Rausch de Traubenberg and
  C.A. Savoy, Phys. Lett. B \textbf{315} (1993) 331, Z. Phys. C {\bf
    67} (1995) 665 and Nucl. Phys. B \textbf{492} (1997) 307;
  U.~Ellwanger, Phys.\ Lett.\  B {\bf 303} (1993) 271; P. Pandita,
  Z. Phys. C \textbf{59} (1993) 575; T. Elliott, S.F. King and
  P.L. White, Phys. Rev. D {\bf 49} (1994) 2435; S.F. King and
  P.L. White, Phys. Rev. D \textbf{52} (1995) 4183;  F.~Franke and
  H.~Fraas, Int.\ J.\ Mod.\ Phys.\  A {\bf 12} (1997) 479. 

\bibitem{Nevzorov:2004ge}
{\it For reviews, see:}
M.~Maniatis,
  Int.\ J.\ Mod.\ Phys.\ {\bf A25} (2010) 3505
  [arXiv:0906.0777 [hep-ph]];
 U.~Ellwanger, C.~Hugonie, A.~M.~Teixeira,
 Phys.\ Rept.\  {\bf 496} (2010) 1
 [arXiv:0910.1785 [hep-ph]];
U.~Ellwanger,
  Eur.\ Phys.\ J.\ C {\bf 71} (2011) 1782
  [arXiv:1108.0157 [hep-ph]].

\bibitem{finetune}
 M.~Bastero-Gil, C.~Hugonie, S.~F.~King, D.~P.~Roy and S.~Vempati,
  Phys.\ Lett.\ B\ {\bf 489}, 359  (2000)
  [hep-ph/0006198];
  A.~Delgado, C.~Kolda, J.~P.~Olson and A.~de la Puente,
  Phys.\ Rev.\ Lett.\ \ {\bf 105}, 091802  (2010)
  [arXiv:1005.1282 [hep-ph]];
U.~Ellwanger, G.~Espitalier-Noel and C.~Hugonie,
  JHEP {\bf 1109} (2011) 105
  [arXiv:1107.2472 [hep-ph]];
  G.~G.~Ross and K.~Schmidt-Hoberg,
  arXiv:1108.1284 [hep-ph];
S.~F.~King, M.~Muhlleitner and R.~Nevzorov,
  Nucl.\ Phys.\ B {\bf 860} (2012) 207
  [arXiv:1201.2671 [hep-ph]];
J.~-J.~Cao, Z.~-X.~Heng, J.~M.~Yang, Y.~-M.~Zhang and J.~-Y.~Zhu,
  JHEP {\bf 1203} (2012) 086
  [arXiv:1202.5821 [hep-ph]];
J.~Cao, Z.~Heng, J.~M.~Yang and J.~Zhu,
  JHEP {\bf 1210} (2012) 079
  [arXiv:1207.3698 [hep-ph]];
S.~F.~King, M.~Muhlleitner, R.~Nevzorov and K.~Walz,
  Nucl.\ Phys.\ B {\bf 870} (2013) 323
  [arXiv:1211.5074 [hep-ph]].

\bibitem{Baglio:2013vya}
  J.~Baglio, T.~N.~Dao, R.~Gr\"ober, M.~M.~M\"uhlleitner, H.~Rzehak, M.~Spira, J.~Streicher and K.~Walz,
  EPJ Web Conf.\  {\bf 49} (2013) 12001.

\bibitem{Ender:2011qh}
K.~Ender, T.~Graf, M.~M\"uhlleitner and H.~Rzehak,
  Phys.\ Rev.\ D {\bf 85} (2012) 075024
  [arXiv:1111.4952 [hep-ph]].

\bibitem{Graf:2012hh}
  T.~Graf, R.~Gr\"ober, M.~M\"uhlleitner, H.~Rzehak and K.~Walz,
  JHEP {\bf 1210} (2012) 122
  [arXiv:1206.6806 [hep-ph]].

\bibitem{effpot}
U.~Ellwanger,
  Phys.\ Lett.\  {\bf B303} (1993)  271 [hep-ph/9302224];
T.~Elliott, S.~F.~King, P.~L.~White,
  Phys.\ Lett.\  {\bf B305} (1993) 71 [hep-ph/9302202],
  Phys.\ Lett.\  {\bf B314} (1993)  56 [hep-ph/9305282],
  Phys.\ Rev.\  {\bf D49} (1994)  2435 [hep-ph/9308309];
P.~N.~Pandita,
  Z.\ Phys.\  {\bf C59} (1993)  575,
  Phys.\ Lett.\  {\bf B318} (1993)  338.

\bibitem{leadlog}
U.~Ellwanger, C.~Hugonie,
  Phys.\ Lett.\  {\bf B623} (2005)  93 [hep-ph/0504269].

\bibitem{Degrassi:2009yq}
G.~Degrassi and P.~Slavich,
  Nucl.\ Phys.\ B {\bf 825} (2010) 119
  [arXiv:0907.4682 [hep-ph]].

\bibitem{full1loop}
F.~Staub, W.~Porod, B.~Herrmann,
  JHEP {\bf 1010} (2010)  040 [arXiv:1007.4049 [hep-ph]].

\bibitem{effcorr1}
S.~W.~Ham, J.~Kim, S.~K.~Oh and D.~Son,
  Phys.\ Rev.\ D {\bf 64} (2001) 035007 [hep-ph/0104144];
S.~W.~Ham, S.~H.~Kim, S.~K.~Oh and D.~Son,
  Phys.\ Rev.\ D {\bf 76} (2007) 115013 [arXiv:0708.2755 [hep-ph]].

\bibitem{effcorr2}
S.~W.~Ham, S.~K.~Oh and D.~Son,
  Phys.\ Rev.\ D {\bf 65} (2002) 075004 [hep-ph/0110052];
S.~W.~Ham, Y.~S.~Jeong and S.~K.~Oh,
  hep-ph/0308264.

\bibitem{effcorr3}
  K.~Funakubo and S.~Tao,
  Prog.\ Theor.\ Phys.\  {\bf 113} (2005) 821
  [hep-ph/0409294].

\bibitem{complex2loop}
K.~Cheung, T.--J.~Hou, J.~S.~Lee and E.~Senaha,
  Phys.\ Rev.\ D {\bf 82} (2010) 075007 [arXiv:1006.1458 [hep-ph]].

\bibitem{Munir:2013dya}
  S.~Munir,
  arXiv:1310.8129 [hep-ph].

\bibitem{Nhung:2013lpa}
  D.~T.~Nhung, M.~M\"uhlleitner, J.~Streicher and K.~Walz,
  JHEP {\bf 11} (2013) 181  [arXiv:1306.3926 [hep-ph]].

\bibitem{hdecay}
  A.~Djouadi, J.~Kalinowski and M.~Spira,
  Comput.\ Phys.\ Commun.\  {\bf 108} (1998) 56
  [hep-ph/9704448];
J.~M.~Butterworth, F.~Maltoni, F.~Moortgat, P.~Richardson, S.~Schumann, P.~Skands, J.~Alwall and A.~Arbey {\it et al.},
  arXiv:1003.1643 [hep-ph];
A.~Djouadi, M.~M.~M\"uhlleitner and M.~Spira,
  Acta Phys.\ Polon.\ B {\bf 38} (2007) 635
  [hep-ph/0609292].

\bibitem{Pierce:1996zz}
R. Hempfling, Phys. Rev. {\bf D49} (1994) 6168; 
L. Hall, R. Rattazzi and U. Sarid, Phys. Rev. {\bf D50} (1994) 7048;
M. Carena, M. Olechowski, S. Pokorski and C.E.M. Wagner, Nucl. Phys. {\bf B426} (1994)
269;
  D.~M.~Pierce, J.~A.~Bagger, K.~T.~Matchev and R.--J.~Zhang,
  Nucl.\ Phys.\ B {\bf 491} (1997) 3
  [hep-ph/9606211];
M.~S.~Carena, S.~Mrenna and C.~E.~M.~Wagner,
  Phys.\ Rev.\ D {\bf 60} (1999) 075010
  [hep-ph/9808312].

\bibitem{Carena:1999py}
  M.~S.~Carena, D.~Garcia, U.~Nierste and C.~E.~M.~Wagner,
  Nucl.\ Phys.\ B {\bf 577} (2000) 88
  [hep-ph/9912516];
M.~S.~Carena, J.~R.~Ellis, S.~Mrenna, A.~Pilaftsis and C.~E.~M.~Wagner,
  Nucl.\ Phys.\ B {\bf 659} (2003) 145
  [hep-ph/0211467].

\bibitem{Guasch:2003cv}
 J.~Guasch, P.~H\"afliger and M.~Spira,
  Phys.\ Rev.\ D {\bf 68} (2003) 115001
  [hep-ph/0305101].

\bibitem{noth1}
  D.~Noth and M.~Spira,
  Phys.\ Rev.\ Lett.\  {\bf 101} (2008) 181801
  [arXiv:0808.0087 [hep-ph]].

\bibitem{noth2}
D.~Noth and M.~Spira,
  JHEP {\bf 1106} (2011) 084
  [arXiv:1001.1935 [hep-ph]].

\bibitem{reisser}
L.~Mihaila and C.~Reisser,
  JHEP {\bf 1008} (2010) 021
  [arXiv:1007.0693 [hep-ph]].

\bibitem{slha1}
P.~Z.~Skands et al.,
  JHEP {\bf 0407} (2004) 036 [hep-ph/0311123].

\bibitem{slha2}
B.~C.~Allanach et al.,
  Comput.\ Phys.\ Commun.\  {\bf 180} (2009) 8
  [arXiv:0801.0045 [hep-ph]].

\bibitem{Frank:2006yh}
  M.~Frank, T.~Hahn, S.~Heinemeyer, W.~Hollik, H.~Rzehak and G.~Weiglein,
  JHEP {\bf 0702} (2007) 047
  [hep-ph/0611326].

\bibitem{Hollik:2002mv} 
  W.~Hollik, E.~Kraus, M.~Roth, C.~Rupp, K.~Sibold and D.~Stockinger,
  Nucl.\ Phys.\ B {\bf 639}, 3 (2002)
  [hep-ph/0204350].

\bibitem{nearthreshold}
E. Braaten and J. P. Leveille, Phys. Rev. D {\bf 22} (1980) 715; 
N. Sakai, Phys. Rev. D {\bf 22} (1980) 2220; 
T. Inami and T. Kubota, Nucl. Phys. B {\bf 179} (1981) 171; 
M. Drees and K.-I. Hikasa, Phys. Rev. D {\bf 41} (1990) 1547; 
M. Drees and K.-I. Hikasa, Phys. Lett. B {\bf 240} (1990) 455
[Erratum-ibid. B {\bf 262} (1991) 497].

\bibitem{abovethreshold}
S. G. Gorishnii, A. L. Kataev, S. A. Larin and L. R. Surguladze, Mod. Phys. Lett. A
{\bf 5} (1990) 2703; 
S. G. Gorishnii, A. L. Kataev, S. A. Larin and L. R. Surguladze,
Phys.Rev. D {\bf 43} (1991) 1633; 
A. L. Kataev and V. T. Kim, Mod. Phys. Lett. A {\bf 9} (1994) 1309;
S. G. Gorishnii, A. L. Kataev and S. A. Larin,
Sov. J. Nucl. Phys. {\bf 40} (1984) 329 [Yad. Fiz. 40 (1984) 517]; 
L. R. Surguladze, Phys. Lett. B {\bf 341} (1994) 60 [hep-ph/9405325];
S. A. Larin, T. van Ritbergen and J. A. M. Vermaseren, Phys. Lett. B
{\bf 362} (1995) 134 [hep-ph/9506465]; 
K. G. Chetyrkin and A. Kwiatkowski, Nucl. Phys. B {\bf 461} (1996) 3
[hep-ph/9505358]. 

\bibitem{chetyrkin}
K. G. Chetyrkin, Phys. Lett. B {\bf 390} (1997) 309 [hep-ph/9608318].

\bibitem{baikov}
P. A. Baikov, K. G. Chetyrkin and J. H. K\"uhn, Phys. Rev. Lett. {\bf
  96} (2006) 012003 [hep-ph/0511063].

\bibitem{Djouadi:1994gf}
A.~Mendez and A.~Pomarol,
  Phys.\ Lett.\ B {\bf 252} (1990) 461;
C.--S.~Li and R.~J.~Oakes,
  Phys.\ Rev.\ D {\bf 43} (1991) 855;
  A.~Djouadi and P.~Gambino,
  Phys.\ Rev.\ D {\bf 51} (1995) 218
   [Erratum-ibid.\ D {\bf 53} (1996) 4111]
  [hep-ph/9406431].

\bibitem{offshellphi}
A. Djouadi, J. Kalinowski and P.M. Zerwas, Z. Phys. C {\bf 70} (1996) 437;
S. Moretti and W.J. Stirling, Phys. Lett. B {\bf 347} (1995) 291 and
(E) B {\bf 366} (1996) 451.

%


\bibitem{Chetyrkin:1997un}
  K.~G.~Chetyrkin, B.~A.~Kniehl and M.~Steinhauser,
  Nucl.\ Phys.\ B {\bf 510} (1998) 61
  [hep-ph/9708255].

\bibitem{Kramer:1996iq}
  M.~Kr\"amer, E.~Laenen and M.~Spira,
  Nucl.\ Phys.\ B {\bf 511} (1998) 523
  [hep-ph/9611272].

\bibitem{Schroder:2005hy}
  Y.~Schr\"oder and M.~Steinhauser,
  JHEP {\bf 0601} (2006) 051
  [hep-ph/0512058].

\bibitem{Chetyrkin:2005ia}
  K.~G.~Chetyrkin, J.~H.~K\"uhn and C.~Sturm,
  Nucl.\ Phys.\ B {\bf 744} (2006) 121
  [hep-ph/0512060].

\bibitem{nloggqcd}
  T.~Inami, T.~Kubota and Y.~Okada,
  Z.\ Phys.\ C {\bf 18} (1983) 69;
  A.~Djouadi, M.~Spira and P.~M.~Zerwas,
  Phys.\ Lett.\ B {\bf 264} (1991) 440;
  K.~G.~Chetyrkin, B.~A.~Kniehl and M.~Steinhauser,
  Phys.\ Rev.\ Lett.\  {\bf 79} (1997) 353
  [hep-ph/9705240].

\bibitem{Spira:1993bb}
 M.~Spira, A.~Djouadi, D.~Graudenz and P.~M.~Zerwas,
 Phys.\ Lett.\ B {\bf 318} (1993) 347.

\bibitem{Spira:1995rr}
 M.~Spira, A.~Djouadi, D.~Graudenz and P.~M.~Zerwas,
 Nucl.\ Phys.\ B {\bf 453} (1995) 17
 [hep-ph/9504378].

\bibitem{Baikov:2006ch}
  P.~A.~Baikov and K.~G.~Chetyrkin,
  Phys.\ Rev.\ Lett.\  {\bf 97} (2006) 061803
  [hep-ph/0604194].

\bibitem{Dawson:1996xz}
  S.~Dawson, A.~Djouadi and M.~Spira,
  Phys.\ Rev.\ Lett.\  {\bf 77} (1996) 16
  [hep-ph/9603423];
A.~Djouadi, V.~Driesen, W.~Hollik and J.~I.~Illana,
  Eur.\ Phys.\ J.\ C {\bf 1} (1998) 149
  [hep-ph/9612362].

\bibitem{Harlander:2003bb}
  R.~V.~Harlander and M.~Steinhauser,
  Phys.\ Lett.\ B {\bf 574} (2003) 258
  [hep-ph/0307346];
R.~Harlander and M.~Steinhauser,
  Phys.\ Rev.\ D {\bf 68} (2003) 111701
  [hep-ph/0308210];
R.~V.~Harlander and F.~Hofmann,
  JHEP {\bf 0603} (2006) 050
  [hep-ph/0507041].

\bibitem{Anastasiou:2006hc}
  C.~Anastasiou, S.~Beerli, S.~Bucherer, A.~Daleo and Z.~Kunszt,
  JHEP {\bf 0701} (2007) 082
  [hep-ph/0611236].

\bibitem{Muhlleitner:2006wx}
  M.~M\"uhlleitner and M.~Spira,
  Nucl.\ Phys.\ B {\bf 790} (2008) 1
  [hep-ph/0612254].

\bibitem{Anastasiou:2008rm}
  C.~Anastasiou, S.~Beerli and A.~Daleo,
  Phys.\ Rev.\ Lett.\  {\bf 100} (2008) 241806
  [arXiv:0803.3065 [hep-ph]].

\bibitem{Degrassi:2008zj}
  G.~Degrassi and P.~Slavich,
  Nucl.\ Phys.\ B {\bf 805} (2008) 267
  [arXiv:0806.1495 [hep-ph]].

\bibitem{Muhlleitner:2008yw}
  M.~M\"uhlleitner, H.~Rzehak and M.~Spira,
  JHEP {\bf 0904} (2009) 023
  [arXiv:0812.3815 [hep-ph]].

\bibitem{Degrassi:2010eu}
  G.~Degrassi and P.~Slavich,
  JHEP {\bf 1011} (2010) 044
  [arXiv:1007.3465 [hep-ph]];
  G.~Degrassi, S.~Di Vita and P.~Slavich,
  JHEP {\bf 1108} (2011) 128
  [arXiv:1107.0914 [hep-ph]];
  G.~Degrassi, S.~Di Vita and P.~Slavich,
  Eur.\ Phys.\ J.\ C {\bf 72} (2012) 2032
  [arXiv:1204.1016 [hep-ph]].

\bibitem{hgagaqcd}
H. Zheng and D. Wu, Phys. Rev. {\bf D42} (1990) 3760;
A. Djouadi, M. Spira, J. van der Bij and P.M. Zerwas, Phys. Lett. {\bf
  B257} (1991) 187;
S. Dawson and R.P. Kauffman, Phys. Rev. {\bf D47} (1993) 1264;
A. Djouadi, M. Spira and P.M. Zerwas, Phys. Lett. {\bf B311} (1993) 255;
K. Melnikov and O. Yakovlev, Phys. Lett. {\bf B312} (1993) 179;
M. Inoue, R. Najima, T. Oka and J. Saito, Mod. Phys. Lett.{\bf A9} (1994)
1189.

\bibitem{Melnikov:1994jb}
K.~Melnikov, M.~Spira and O.~I.~Yakovlev,
Z.\ Phys.\ C {\bf 64} (1994) 401
[hep-ph/9405301].

\bibitem{nloZga}
  M.~Spira, A.~Djouadi and P.~M.~Zerwas,
  Phys.\ Lett.\ B {\bf 276} (1992) 350.

\bibitem{cahn}
R.N. Cahn, Rep. Prog. Phys. {\bf 52} (1989) 389.

\bibitem{Miller:2003ay}
{\it For a comprehensive study of the NMSSM Higgs masses, see:}
  D.~J.~Miller, 2, R.~Nevzorov and P.~M.~Zerwas,
  Nucl.\ Phys.\ B {\bf 681} (2004) 3
  [hep-ph/0304049].

\bibitem{fortsch}
M.~Spira,
  Fortsch.\ Phys.\  {\bf 46} (1998) 203
  [hep-ph/9705337];
A.~Djouadi,
  Phys.\ Rept.\  {\bf 459} (2008) 1
  [hep-ph/0503173].

\bibitem{Djouadi:1996pj}
  A.~Djouadi, J.~Kalinowski, P.~Ohmann and P.~M.~Zerwas,
  Z.\ Phys.\ C {\bf 74} (1997) 93
  [hep-ph/9605339].

\bibitem{Djouadi:1996mj}
  A.~Djouadi, P.~Janot, J.~Kalinowski and P.~M.~Zerwas,
  Phys.\ Lett.\ B {\bf 376} (1996) 220
  [hep-ph/9603368].

\bibitem{Bartl:1997yd}
  A.~Bartl, H.~Eberl, K.~Hidaka, T.~Kon, W.~Majerotto and Y.~Yamada,
  Phys.\ Lett.\ B {\bf 402} (1997) 303
  [hep-ph/9701398];
 A.~Arhrib, A.~Djouadi, W.~Hollik and C.~Junger,
  Phys.\ Rev.\ D {\bf 57} (1998) 5860
  [hep-ph/9702426].

\bibitem{Eberl:1999he}
 H.~Eberl, K.~Hidaka, S.~Kraml, W.~Majerotto and Y.~Yamada,
 Phys.\ Rev.\ D {\bf 62} (2000) 055006
 [hep-ph/9912463].

\bibitem{Accomando:2011jy}
  E.~Accomando, G.~Chachamis, F.~Fugel, M.~Spira and M.~Walser,
  Phys.\ Rev.\ D {\bf 85} (2012) 015004
  [arXiv:1103.4283 [hep-ph]].

\bibitem{loggres}
J. Ellis, M.K. Gaillard and D.V. Nanopoulos, Nucl. Phys. {\bf B106}
(1976) 292;
H. Georgi, S. Glashow, M. Machacek and D.V. Nanopoulos,
Phys. Rev. Lett. {\bf 40} (1978) 692.

\bibitem{vainshteineal}
A.I. Vainshtein, M.B. Voloshin, V.I. Zakharov and M.A. Shifman,
Sov. J. Nucl. Phys. {\bf 30} (1979) 711.

\bibitem{coarasa}
J.A. Coarasa, R.A. Jim\'enez and J. Sol\`a, Phys. Lett. B {\bf 389}
(1996) 312.

\bibitem{Contino:2013kra}
  R.~Contino, M.~Ghezzi, C.~Grojean, M.~M\"uhlleitner and M.~Spira,
  JHEP {\bf 1307} (2013) 035
  [arXiv:1303.3876 [hep-ph]].

\end{thebibliography}
\end{document}